\documentclass[article, twocolumn, notitlepage,superscriptaddress, longbibliography]{revtex4-2}
\usepackage{natbib}
\usepackage{graphicx}
\usepackage{amssymb}
\usepackage{amsmath}
\usepackage{ifthen}
\usepackage{braket}
\usepackage{xcolor}
\usepackage{bm}
\usepackage{bbm}
\usepackage{bbm}
\usepackage{maybemath}
\usepackage{comment}
\usepackage{siunitx}
\usepackage{float}
\usepackage{dcolumn}
\newcolumntype{d}[1]{D{.}{.}{#1}}
\usepackage{subfigure}

\definecolor{garrosgreen}{rgb}{0.1, 0.4, 0.1}
\definecolor{dartmouthgreen}{rgb}{0.05, 0.5, 0.06}
\definecolor{duelferred}{rgb}{0.7, 0.2, 0.1}
\definecolor{cambridgeblue}{rgb}{0.1, 0.3, 1.0}
\definecolor{oxfordblue}{rgb}{0.05, 0.2, 0.7}

\newcommand{\addrFSU}{Department of Chemistry, Physics and Materials Science,
Fayetteville State University, Fayetteville, NC 28301, USA}

\newcommand{\addrSRNL}{Savannah River National Laboratory, Aiken, SC 29808, USA}

\newcommand{\addrNCAT}{Department of Chemistry, North Carolina A \& T State University, Greensboro, NC, 27411, USA}

\newcommand{\addrNDSU}{Department of Chemistry and Biochemistry, North Dakota State University, Fargo, ND 58108, USA}

\begin{document}

\title{Confinement of quasi-atomic structures in Ti$_{\mathbf{2}}$N and Ti$_{\mathbf{3}}$N$_{\mathbf{2}}$  MXene Electrides}

\author{Chandra M. Adhikari}
\affiliation{\addrFSU}

\author{Dinesh Thapa}
\affiliation{\addrNDSU}

\author{Talon D. Alexander}
\affiliation{\addrFSU}

\author{Christopher K. Addaman}
\affiliation{\addrFSU}

\author{Shubo Han}
\affiliation{\addrFSU}

\author{Bishnu P. Bastakoti}
\affiliation{\addrNCAT}

\author{Daniel E. Autrey}
\affiliation{\addrFSU}

\author{Svetlana Kilina}
\affiliation{\addrNDSU}

\author{Binod K. Rai}
\affiliation{\addrSRNL}

\author{Bhoj R. Gautam}
\affiliation{\addrFSU}

\begin{abstract}
Metal carbides, nitrides, or carbonitrides of early transition metals, better known as MXenes, possess notable structural, electrical, and magnetic properties. Analyzing electronic structures by calculating structural stability, band structure, density of states, Bader charge transfer, and work functions utilizing first principle calculations, we revealed that titanium nitride Mxenes, namely Ti$_2$N and Ti$_3$N$_2$,  have excess anionic electrons in their pseudo-atomic structure inside the crystal lattice, making them MXene electrides.  Bulk Ti$_3$N$_2$ has competing antiferromagnetic (AFM) and ferromagnetic(FM) configurations with slightly more stable AFM configurations, while the Ti$_2$N MXene is nonmagnetic. Although Ti$_3$N$_2$ favors AFM configurations with hexagonal crystal systems having $6/mmm$ point group symmetry, Ti$_3$N$_2$ does not support altermagnetism. The monolayer of the Ti$_3$N$_2$ MXene is a ferromagnetic electride. These unique properties of having non-nuclear interstitial anionic electrons in the electronic structure of titanium nitride MXene have not yet been reported in the literature.  Density functional theory calculations show TiN is neither an electride, MXene, or magnetic.
\end{abstract}

\maketitle


\section{\label{sec:level1} Introduction}

Newly discovered two-dimensional layered inorganic early transition metal carbides, nitrides, and carbonitrides compounds with the general formula of M$_{n+1}$X$_{n}$, where M is a transition metal found in groups  IIIB--VIB and periods 4-6 and X being carbon (C) and/or nitrogen (N), have received significant attention in the materials science community because of their very useful properties required in various applications within the electronic, magnetic and optical industries~\cite{Naguib_adma_2021}. The name MXenes is adopted for M$_{n+1}$X$_{n}$ compounds to emphasize their composition and similarities with the well-known two-dimensional (2D) material graphene~\cite{Naguib_adma_2011}. 

Since they are layered 2D structures, MXenes have a larger surface-to-volume ratio, similar to graphene. Mxenes have high electrical conductivity and a larger oxidation-reduction reaction active surface, which improves electrode-electrolyte ion exchange and heat dissipation, making them highly efficient materials in the industry of rechargeable batteries and supercapacitors~\cite{BILIBANA_AdvSenEngMat_2023, Gogotsi_acsnano_2019}.  MXenes have a tunable dielectric response, surface-enhanced Raman scattering, high light-to-heat conversion efficiency, and plasmonic broadband absorption, which make them a good candidate for opto-plasmonics~\cite{Chaudhuri_acsphotonics_2018, Soundiraraju_acsnano_2017}. MXenes have biomedical applications as antibacterial agents, biosensing, therapeutic diagnosis, implants, and bioimaging~\cite{Li_JNano_2023}. MXenes provide electromagnetic interference shielding, even at a minimal thickness, due to their high electrical conductivity and multiple internal reflections from their flakes~\cite{Shahzad_Science_2024}. MXene-enabled technologies work as an environmental cleaning agent, removing organic and inorganic pollutants present in the environment. They also help to recycle and recover precious rare earth elements, eliminating radionuclides and heavy ions such as Cr$^{6+}$, Cd$^{2+}$, and Pb$^{2+}$ from electronic wastes~\cite{Bury_MRS_2023}. Tunable bandgap with proper selection of surface termination group, superior surface properties, and high mechanical strength make MXenes excellent sensing materials for strain/stress, gas, electrochemical, optical, and humidity sensors~\cite{Pei_ACSNano_2021}. MXenes can also be used in catalysts for hydrogen evolution reactions (HER),  oxygen evolution reactions (OER), and bifunctional with HER at the cathode and OER at the anode due to their metallic and intrinsic hydrophilic nature~\cite{Zubair_IJHydEng_2022, Djire_acscatal_2021, Johnson_nonoscale_2022}. 

To understand the MXenes' usability in building magnetic sensors, magnetic storage devices, and in spintronics, one must investigate the magnetism of various MXenes. The magnetic properties prediction of MXenes is quite complicated as they depend not only on the parent early transition metals (M) and carbon and/or nitrogen (X)  of MXenes but also on many other things such as route of synthesis, strain level, purity of the sample, defect, doping, crystal symmetry, oxidation state of M, and the presence of surface termination groups~\cite{Allen_Materials_2021, Frey_Springer_2019, Limbu_J.Magn.Magn.Mats_2022, Sakhraoui_acsomega_2022}. 

Interestingly, since the discovery of MXenes in 2011 at Drexel University~\cite{Naguib_adma_2011}, most studies on MXenes are concentrated more towards carbides than nitrides. With the currently available technology, carbide MXenes are comparatively easier to synthesize and characterize than nitride MXenes. As far as technological applications are concerned, a study showed that nitride-based MXenes have a comparatively larger active surface, a stronger preference to adhere surface termination groups, and higher electrical conductivity than carbide-based Mxenes ~\cite{Zhang_2DMaterials_2018}. To further deepen the knowledge and unearth the fundamental understanding of nitride-based MXenes, we devote this work to titanium nitride MXenes, specifically Ti$_2$N and Ti$_3$N$_2$. A study on TiN is also presented for comparison purposes. Here and onwards, TiN, Ti$_2$N, and Ti$_3$N$_2$ will be collectively referred to as Ti-N compounds.

To better analyze the electronic structure and magnetism of MXenes, a discussion on transition metal to carbon or nitrogen bonding is necessary. Transition metals have comparatively lower electronegativity than carbon and nitrogen. For example, the electronegativity of Titanium (Ti) by the Pauling scale is 1.54, while the same for carbon (C) and nitrogen (N) are 2.55 and 3.04, respectively~\cite{Pauling_nature_1960, Wolfram_Online_2023}. The absolute electronegativity differences of (Ti, C) and (Ti, N) pairs are less than 1.7 but more than 0.4. From a simple rule of thumb of bond classification~\cite{Nickelson_Springer_2019},  both carbide and nitride Mxenes of titanium have neither a pure ionic bonding nor a covalent bonding but prefer to have polar covalent bonds, while the covalent bonding percentage is higher in the Ti to C bond than the Ti to N bond. The electronic charge distribution in Ti$_3$N$_2$ and Ti$_2$N Mxenes is complex and does not follow the bond classification rules as they have unique electronic structures.  Our first principle DFT calculations demonstrate that both Ti$_3$N$_2$ and Ti$_2$N have excess electrons in their pseudo-atomic structure inside their crystal lattice, known as the quasi-Farbe color centers, making them the first discovered titanium-based MXene electrides.  In contrast, no quasi-Farbe color centers are present in TiN crystals, dismissing the possibility of TiN being electride.

An introduction to electride materials, a relatively under-explored material class,  is in order. Electrides have fully trapped isolated electrons occupying interstitial space between atoms with an anisotropic electronic distribution, which are \textit{(i)} neither fully transferred as in ionic, \textit{(ii)} nor shared equally between elements with almost the same electronegativity as in covalent,  \textit{(iii)} nor shared unequally between element with slightly different electronegativity as in polar covalent compounds, but instead, they have pseudo-atomic structures also called empty spheres of interstitial ionic electron (IAE) sites. As these empty spheres have no nucleus, the trapped anionic electrons in IAEs can easily diffuse, resulting in low work functions and high electrical conductivity~\cite{Liu_JMatsChemC_2020}, which makes them a potential candidate for field emission display in cold cathodes, thermionic emitters, battery anode in high-performance energy storage systems, supercapacitors, high-speed transistors and so on~\cite{Huang_ChemPhysLett_1990, TANG_CeramicsInt_2021, Hosono_Philos_2015, Hu_ACSApplMaterInterfaces_2015}. High electron density, high electron mobility, and the ability to activate nitrogen molecules make electrides excellent catalysts for organic molecule synthesis~\cite{Ye_GreenChem_2017}. Exploiting the electrons' spin and charge distribution,  magnetic electrides can be used in spintronic devices for information processing and storage~\cite{Zhang_JAmChemSoc_2023}.  


Carrying out high-throughput material screening in conjunction with a geometry-based algorithm on a large set of materials from the Materials Project database~\cite{Jain_APLM_2013}, Zhou et al. reconfirmed already reported 12 electrides, namely, ScCl~\cite{wan_npj_2018}, YCl~\cite{wan_npj_2018}, ZrCl~\cite{Zhang_PRX_2017}, ZrBr~\cite{Daake_InoChem_1977},  Ca$_2$N~\cite{Lee_nature_2013, Walsh_JMaterChemC_2013}, Sr$_2$N~\cite{Walsh_JMaterChemC_2013}, Ba$_2$N~\cite{Walsh_JMaterChemC_2013}, Sc$_2$C~\cite{McRae_JAmsCheSoc_2022}, Y$_2$C~\cite{Zhang_ChemMat_2014}, Tb$_2$C~\cite{Novoselov_jpcc_2021}, Dy$_2$C~\cite{Novoselov_jpcc_2021}, and Ho$_2$C~\cite{Novoselov_jpcc_2021} and uncovered  12 new materials as layered electrides, namely, TbCl, TbBr, LaBr$_2$, ThI$_2$, La$_2$Br$_5$, Y$_2$Cl$_3$,
Gd$_2$Br$_3$, Sc$_5$Cl$_8$, Tb$_5$Br$_8$,  Er$_6$I$_7$, Sc$_7$Cl$_{10}$, and Ba$_2$LiN~\cite{Zhou_chemmater_2019}. Utilizing high throughput computational screening followed by the first principle calculations, Thapa et al. recently reported that electride material need not necessarily be carbide, nitride, or halide, but they can be a transition metal oxide or chalcogenides such as Zr$_2$X, where X= O, Se, and Te~\cite{Thapa_MaterHoriz_2024}.

Considering both rare earth scandium (Sc) and yttrium (Y) as early transition metal candidates for MXenes, only two carbide MXenes, Sc$_2$C and Y$_2$C, are reported as electrides in the literature~\cite{McRae_JAmsCheSoc_2022, Zhang_ChemMat_2014}. 
A layered rare earth lanthanide gadolinium carbide, Gd$_2$C, is also reported to have a two-dimensional electron gas, whose excess anionic electron counts increase with the square of the electric field intensity~\cite{Chae_PRB_2021}.
The electronegativity of metal cations contributes to determining the electronic structure of electrides. For example, less electronegative yttrium can not open a band gap in Y$_2$C and remains a semi-metallic electride. In contrast, high electronegative scandium and aluminum in Sc$_2$C and Al$_2$C with the same structure as Y$_2$C open up indirect band gaps. The gap is larger for Al$_2$C than Sc$_2$C, implying that the band gap increases with an increase in electronegativivity~\cite{McRae_JAmsCheSoc_2022}.
We here make an extensive examination via the first principle DFT calculations and report a new electride material family:  Ti$_{n+1}$N$_n$ MXene.  The first principle DFT calculations confirm the presence of IAEs in Ti$_3$N$_2$ and Ti$_2$N, but not in TiN.  A few additional benefits of the MXene electride family that stand out from other existing electrides are their thickness tunability due to being a stacked layered structure with $c/a > 1$, higher charge density, and enhanced charge carrier mobility. 

\begin{table}[ht]
\small
  \caption{\  Lattice parameters for computationally optimized structures of  TiN, Ti$_2$N, Ti$_3$N$_2$ FM and Ti$_3$N$_2$ AFM. The values in parenthesis are experimental values presented in Ref.~\cite{Brager_ActaP_1939} for TiN and Ref.~\cite{ Lobier_Comptes_1969} for Ti$_2$N. To the best of our knowledge, no experimental values for lattice parameters are yet available for Ti$_3$N$_2$.}
  \label{tbl:LP}
  \begin{tabular*}{0.48\textwidth}{@{\extracolsep{\fill}}llll}
    \hline
compounds &$a$ (\AA)&$b$ (\AA)& $c$ (\AA) \\
\hline
TiN& 4.241 (4.244)  & 4.241(4.244) & 4.241(4.244) \\ 
Ti$_2$N& 4.149 (4.140) & 4.149 (4.140) & 8.873(8.805) \\
Ti$_3$N$_2$ FM& 3.032 & 3.032 & 14.630 \\
Ti$_3$N$_2$ AFM& 3.032 & 3.032 & 14.629  \\
    \hline
  \end{tabular*}
\end{table}

\section{\label{sec:level2}Structural analysis}

\begin{figure}[htb!]
\includegraphics[scale=0.10, width=0.91\linewidth]{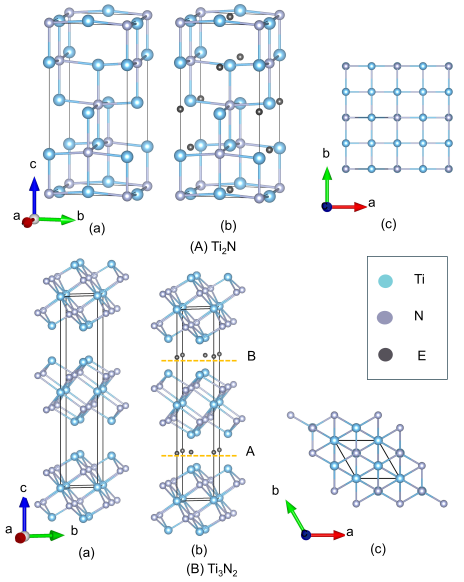}
\caption{\label{fig:structure} Structure models of  Ti$_2$N (A), and Ti$_3$N$_2$ (B) showing unit cells along the rotated (100) axis to a side-view without empty spheres (a) and with empty spheres (b) and a top-view along the (001) crystal axis (c). Empty spheres in two different interlayer regions in Ti$_3$N$_2$ receive different environments. Thus, we denote them as $A$ and $B$ empty spheres for A and B empty sphere layers.   Empty spheres in  Ti$_2$N receive an indifferent crystallographic environment.}
\end{figure}
%

%
\begin{figure*}[htb!]
\includegraphics[width=0.335\linewidth]{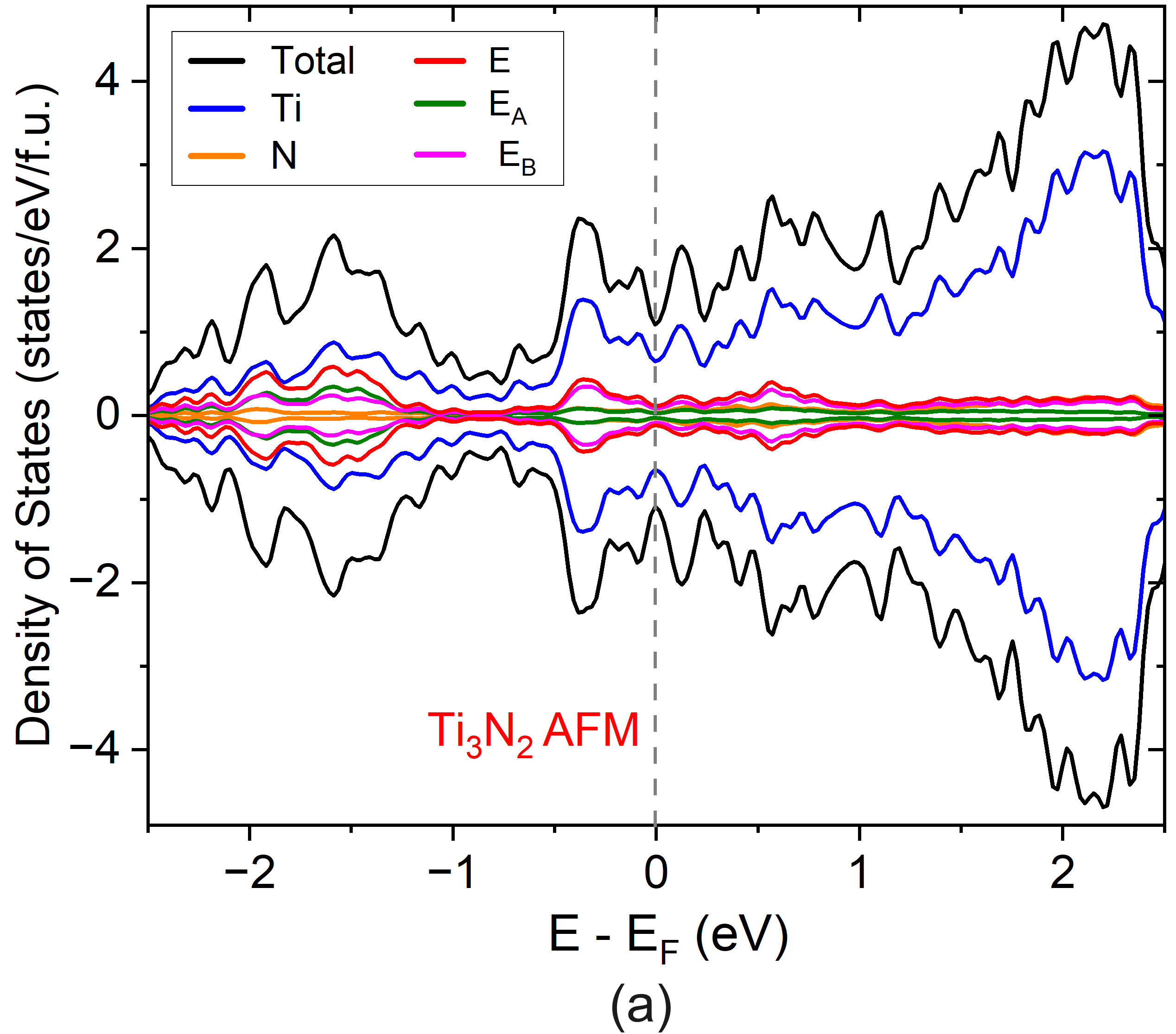}
\includegraphics[width=0.325\linewidth]{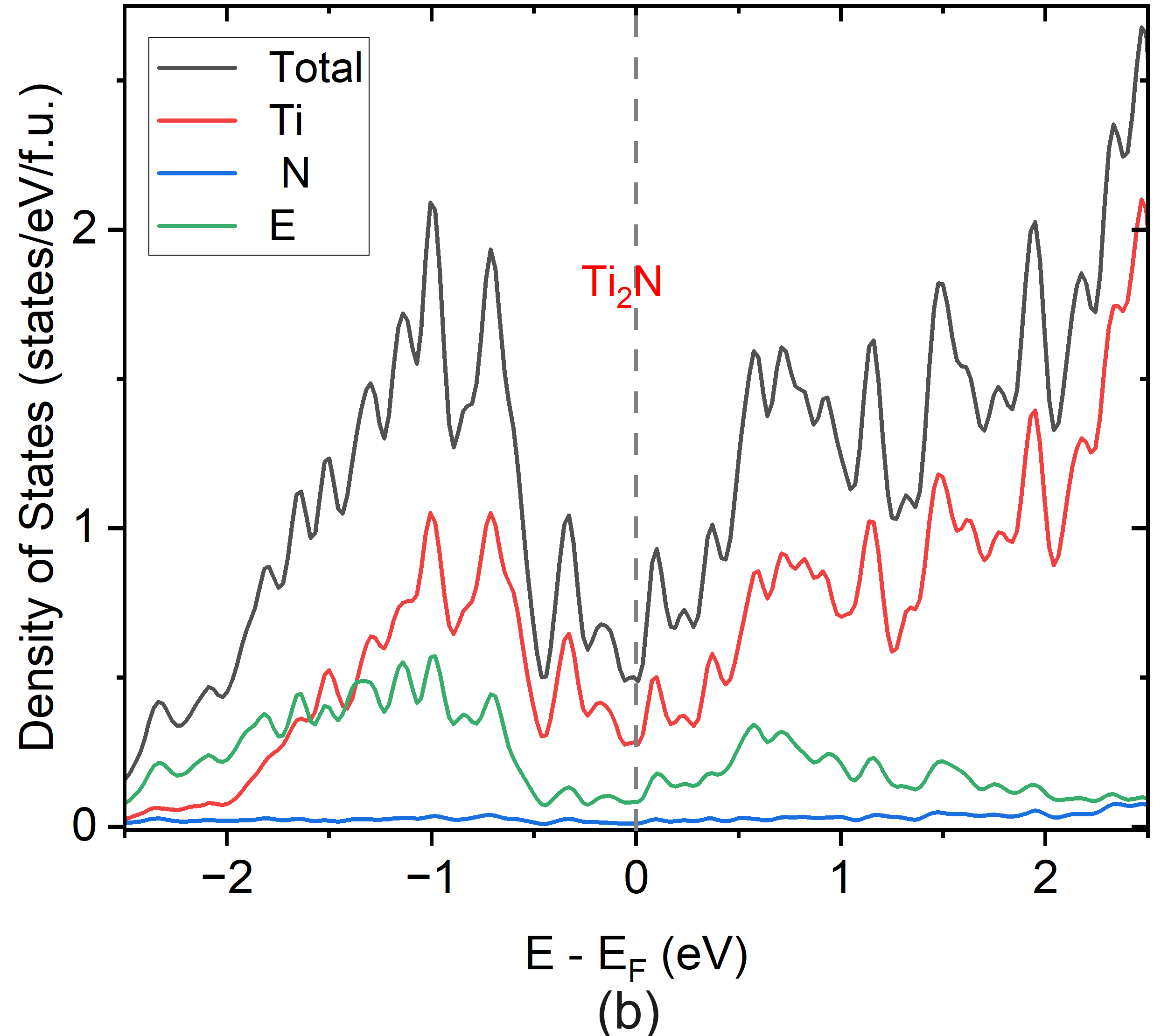}
\includegraphics[width=0.325\linewidth]{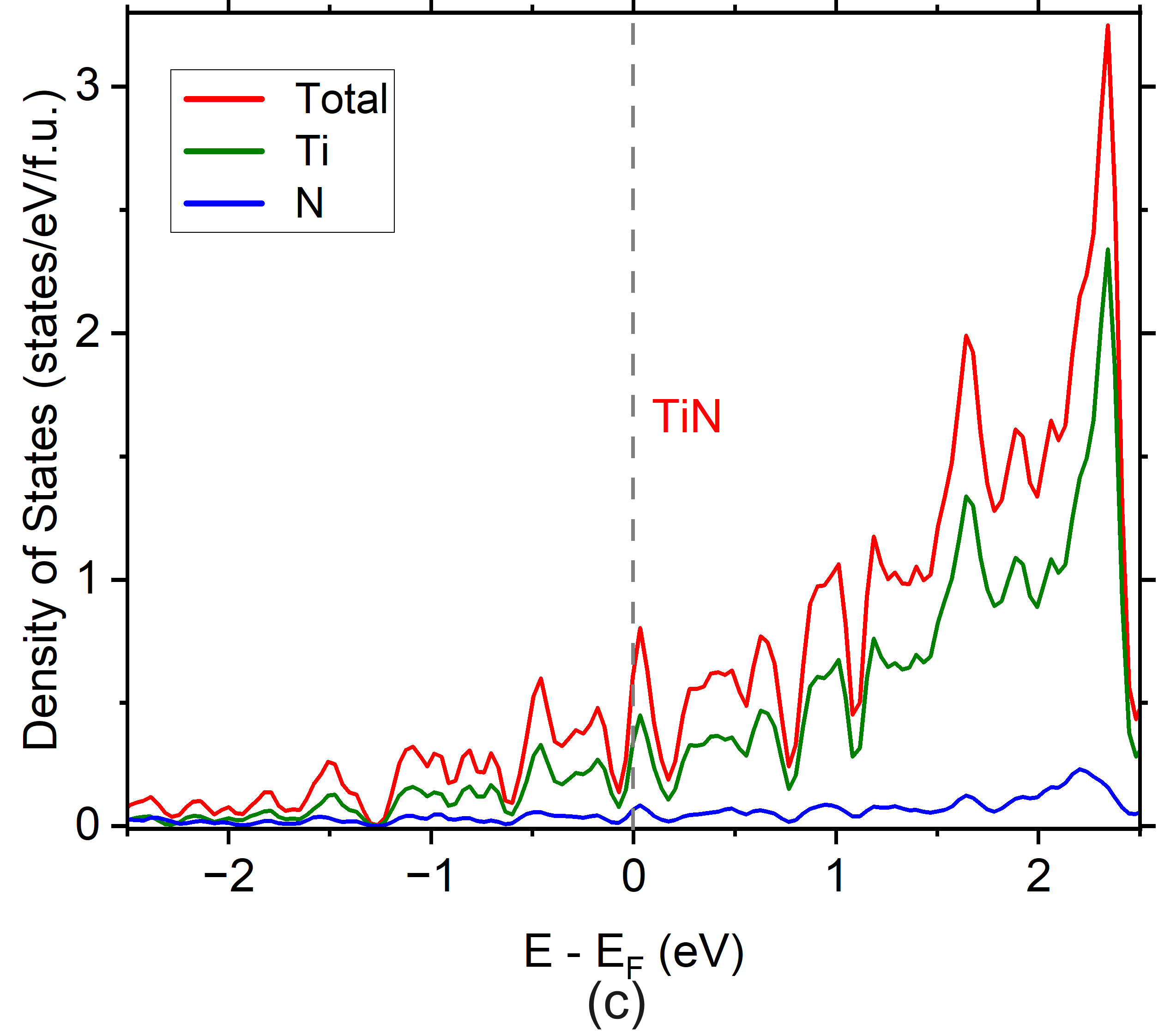}
\caption{\label{fig:structure:Dos} Atom resolved projected density of states of AFM Ti$_3$N$_2$ in bulk phase (a), Ti$_2$N (b), and TiN (c) per their formula unit (f.u.).E$_A$ and E$_B$ represent  DOS of $A$ and $B$ empty spheres per f.u. whereas (E=E$_A$ +E$_B$ )  is the total DOS of empty spheres in the respective crystal structure. Calculation shows a vanishing empty sphere DOS for TiN, indicating that TiN is a nonelectride. }
\end{figure*}
%

%
\begin{figure*}[htb!]
\includegraphics[width=0.335\linewidth]{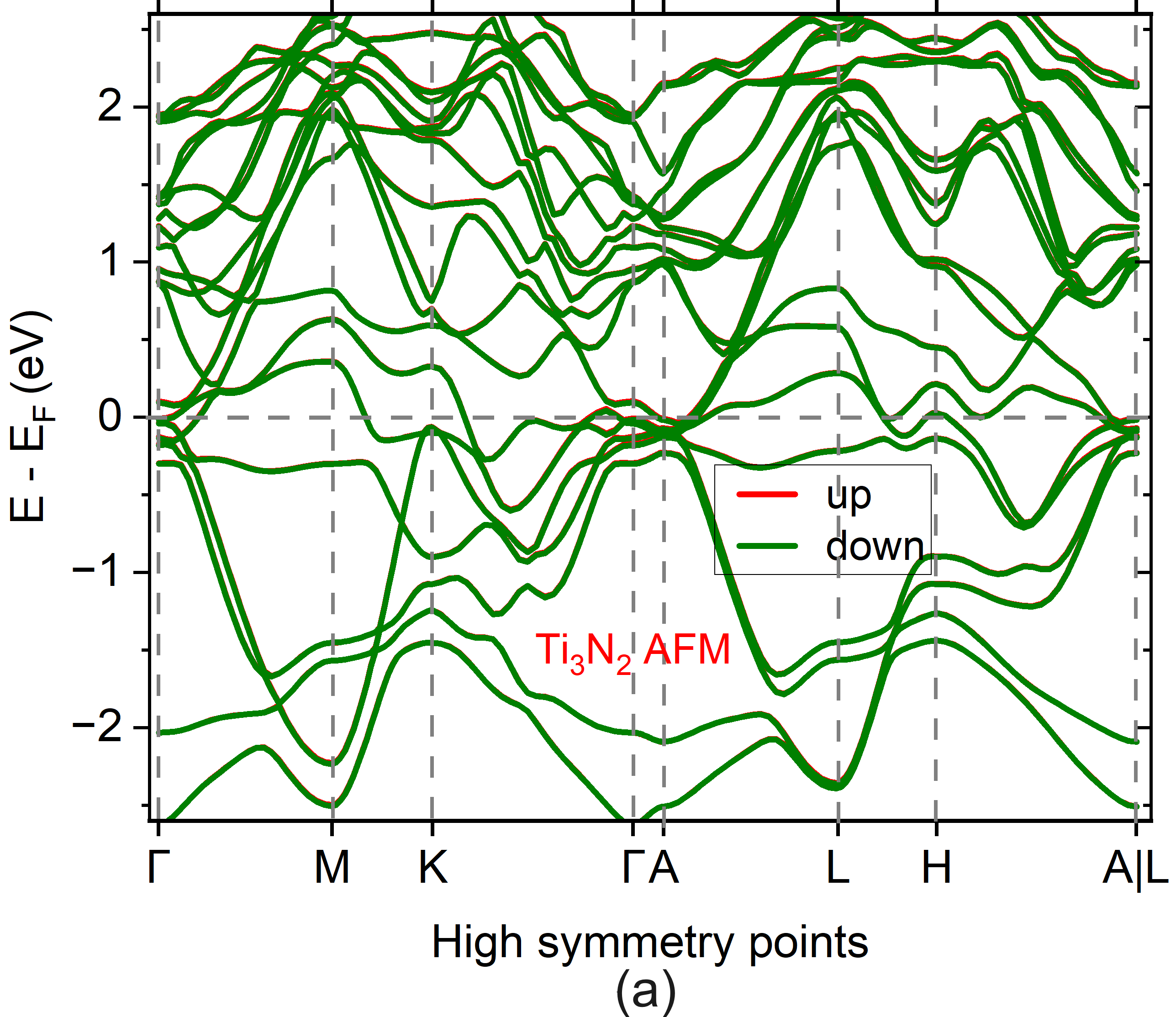}
\includegraphics[width=0.325\linewidth]{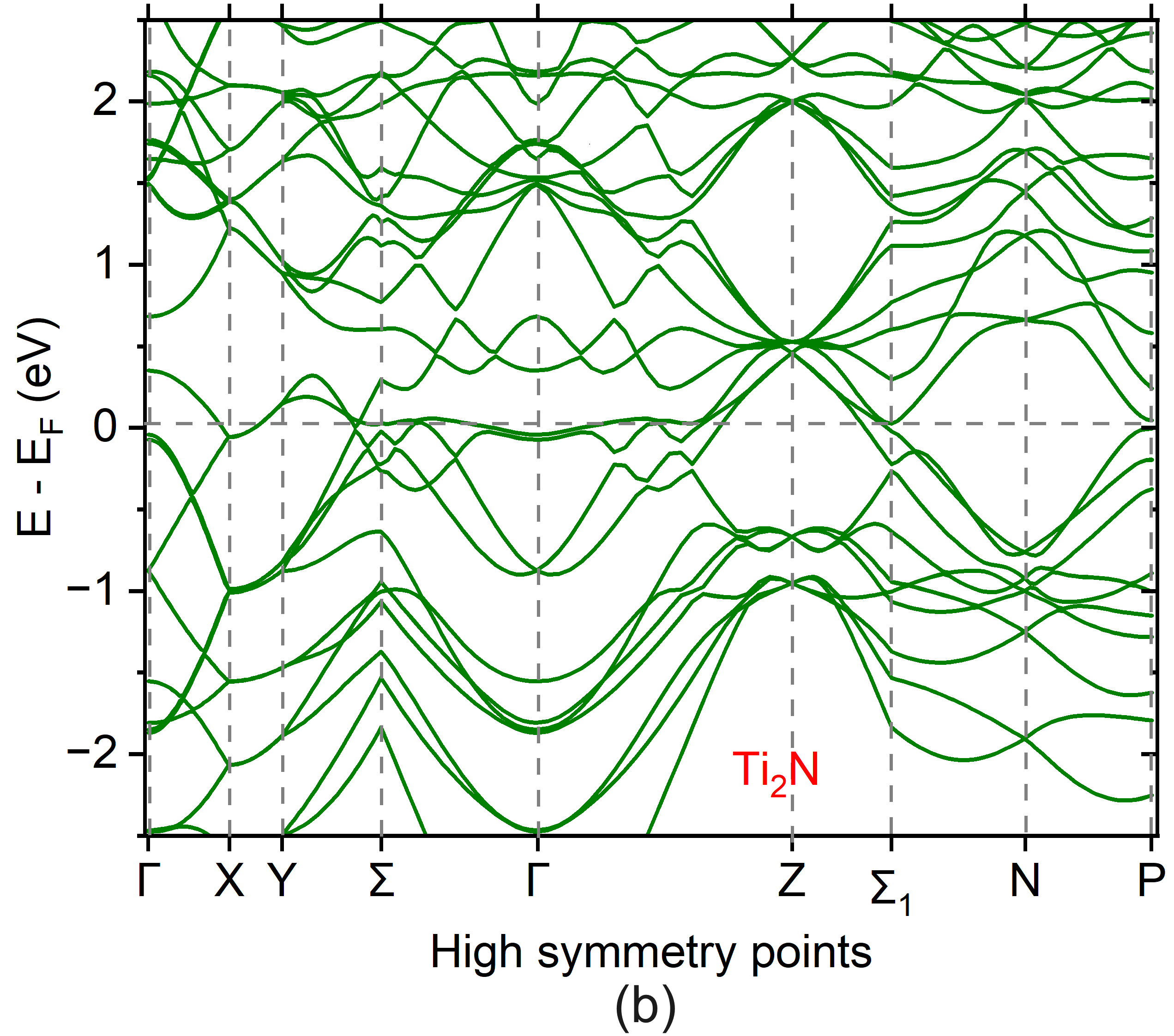}
\includegraphics[width=0.325\linewidth]{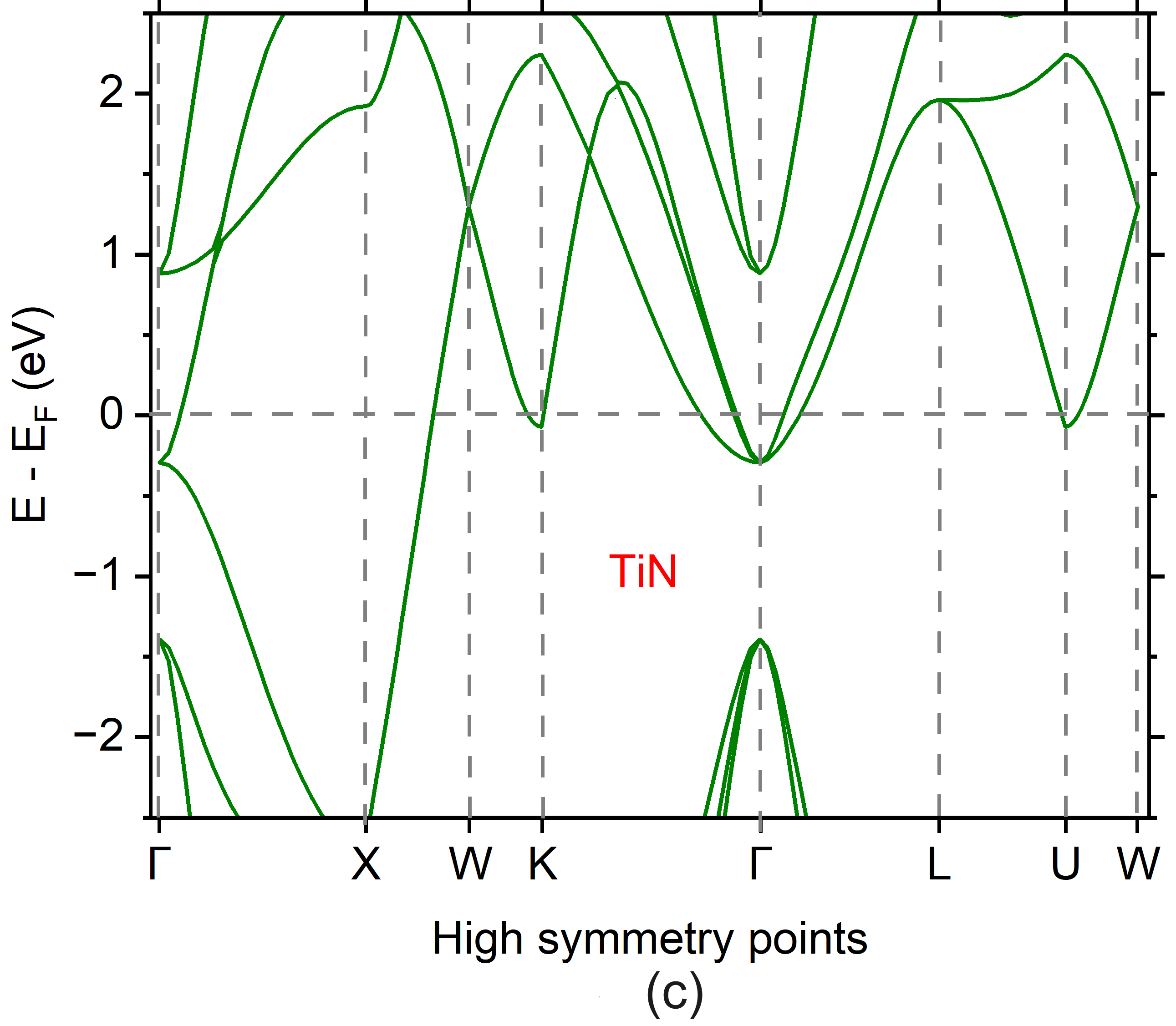}
\caption{\label{fig:structure:BNDS} Bandstructure of AFM Ti$_3$N$_2$ (a), Ti$_2$N (b), and TiN (c) along their respective high symmetry points. Fermi level (E$_{\rm F}$) is shifted to zero energy level. Each of the titanium nitrides is gapless, indicating that they are metallic.  }
\end{figure*}

Ti$_3$N$_2$ forms a hexagonal crystal (p6$_3$/mmc space group) with a $c/a$ ratio of 4.8244 and a $c$ value of 14.6285\,\AA. 
Ti$_2$N crystallizes into a tetragonal structure (I4$_1$/amd space group) with a $c/a$ ratio of 2.1387 and a $c$ value of 8.8725\,\AA.  Ti$_3$N$_2$ and Ti$_2$N have layered structures and experimentally feasible c-values to synthesize 2D-MXenes. 
The titanium in Ti$_3$N$_2$ at the $2a$ Wyckoff position cages itself within a TiN$_6$ octahedron by bonding to six equivalent nitrogen atoms, while the titanium at the $4f$ Wyckoff position forms a triangular noncoplanar-environment (not shown in the figure). The contentious symmetry measures show that titanium in the $2a$ Wyckoff position has less distortion in the coordination environment than titanium in the $4f$ Wyckoff position. Nitrogens in Ti$_3$N$_2$ are at the $4f$ Wyckoff position and confined at the center of NTi$_6$ octahedron.  
In Ti$_2$N, each Ti resides in the $8e$ Wyckoff position and is bonded to three N atoms in a plane. The Nitrogen in Ti$_2$N is at the $4a$ Wyckoff position and bonds to six Ti, forming an NTi$_6$ octahedron.  We have also analyzed TiN for comparison purposes, although it is not an MXene. TiN crystallizes into cubic (Fm$\overline{3}$m space group) with vanishing continuous symmetry measures indicating no distortion in its coordination environment. In TiN, titanium resides in the $4a$ Wyckoff position and nitrogen at the $4b$ Wyckoff position.  In TiN, each titanium and nitrogen are caged in a TiN$_6$ and  NTi$_6$ octahedron. 
Lattice parameters for computationally optimized structures of  TiN, Ti$_2$N, Ti$_3$N$_2$ FM, and Ti$_3$N$_2$ AFM and available relevant experimental values are presented in Table~\ref{tbl:LP}. As per the need for comparison, one can convert the lattice parameter in a face-centered cubic (FCC) to that of corresponding body-centered cubic (BCC) structures using $\ell_{\rm FCC} = \sqrt{2} \; \ell_{\rm BCC}$, where $\ell$ denotes any of the $a$, $b$ or $c$ lattice parameters. 

To measure the stability of any structure, one may need to analyze the formation enthalpy ($E_{F}$) and cohesive energy  ($E_{C}$) of the system, where the former is the energy required to form a crystal from its constituent elements in their most stable bulk state while the latter represents the energy to assemble isolated constituent atoms to a crystal or the same to disassemble a crystal into its constituent isolated atoms. More implicitly, $E_{F}$ is the total energy of the optimized crystal structure minus the bulk energies of the crystal structures of all constituent atoms with their respective weight, and $E_{C}$ is the total energy of the optimized crystal structure minus the sum of the energies of all the isolated atoms. For example, for Ti$_{n+1}$N$_n$ MXenes, the enthalpy of formation E$_F$(Ti$_{n+1}$N$_n$) and the cohesive energy  E$_C$(Ti$_{n+1}$N$_n$ ) read as follows: 
\begin{align}\label{E:formation}
E_{F/C} (\rm{Ti}_{n+1} {N}_n ) &= E_T (\rm{Ti}_{n+1} {\rm N}_n )\nonumber\\
&-(n+1)\, E_{B/I} ( \rm{Ti})-n\,E_{B/I} (\rm{N}) \,,    
\end{align}
where $E_T$ is the total energy of specified MXene,  E$_B$(Ti), and E$_I$(Ti) represent the energy of the bulk titanium metal and the energy of an isolated titanium metal atom. A similar description holds for nitrogen. One may want to divide the right side of the expression by $2n+1$ to obtain the corresponding energy per atom.
For TiN, Eq.~\eqref{E:formation} reads 
\begin{align}\label{E:formation:TiN}
E_{F/C} (\rm{Ti N}) &= E_T (\rm{Ti N} )- E_{B/I} ( \rm{Ti})-\,E_{B/I} (\rm{N})\,. 
\end{align}
Formation enthalpy for Ti$_2$N is $-1.3165$\,eV/atom. The formation enthalpy for antiferromagnetic (AFM) and ferromagnetic (FM) configurations of Ti$_3$N$_2$ are very close to each other, which differs only in the fifth significant figure when expressed in the units of eV/atom. More explicitly,  $E_{F,{\rm AFM}} (\rm{Ti}_{3} {N}_2 ) = -1.3589\,$eV/atom and $E_{F,{\rm FM}} (\rm{Ti}_{3} {N}_2 ) = -1.3583\,$eV/atom, making  AFM configuration slightly more favorable. The formation energy of TiN crystal is $-1.7386\,$eV/atom, making it the most stable among the three titanium nitrides discussed in this article. The formation enthalpy is negative for each titanium nitride of interest in this paper, signifying that the formation of titanium nitride crystals is exothermic and energetically favorable. The cohesive energy per atom for TiN, Ti$_2$N, Ti$_3$N$_2$ AFM, and Ti$_3$N$_2$ FM configurations are $-4.4561$, $-4.9430$, $-4.6218$, and $-4.6212$\,eV respectively.

To facilitate efficient and scalable production of MXenes, in addition to the formation of enthalpy and cohesive energy, one needs to investigate the energy required to exfoliate MXenes from their respective MAX phase, called exfoliation energy. Pristine MXenes are usually exfoliated from Ti$_{n+1}$AX$_n$, where the A element is usually aluminum or gallium. The exfoliation energy  for constructing  Ti$_{n+1}$N$_n$  by removing the A layer is calculated using the following formula~\cite{Jung_nanolett_2018, Yorulmaz_JPhysEng_2020}: 
\begin{align}
E_{\rm exf}=\frac{1}{2\mathcal{A}}\left[2E (\rm A)+ 2E({\rm Ti}_{n+1} {\rm N}_n )-E({\rm Ti}_{n+1} \rm{ A N}_n )\right]\,,
\end{align}
where E(A) is the total energy of an A atom in its preferred bulk configuration. For the aluminum (Al) atom,  cubic $F\overline{m}3m$ (space group \# 225) is the preferred crystal structure with a total energy of  -3.7478\,eV per atom.  Similarly, E(Ti$_{n+1}$N$_n$) and  E(Ti$_{n+1}$AN$_n$ ) are total energy values for Ti$_{n+1}$X$_n$ MXene and Ti$_{n+1}$AX$_n$ MAX phase respectively. $\mathcal{A}$ is the area of the exfoliated layer. The calculated exfoliation energy for Ti$_2$N and Ti$_3$N$_2$  are  0.0224 and 0.0769 eV/\AA$^2$ indicating that Ti$_2$N can be exfoliated more easily than Ti$_3$N$_2$ from their respective Al-based MAX phases.

\section{\label{sec:level3}Computational methods}

The first-principles calculations were performed based on Density Functional Theory (DFT) using the Vienna Ab initio Simulation Package (VASP)~\cite{Kresse_ComMatSci_1996, Kresse_PRB_1996}. The ion-electron interactions were accounted for using the projector augmented wave method~\cite{Blochl_PRB_1994}, and exchange-correlation potentials were approximated by the Perdew–Burke–Ernzerhof functional~\cite{Perdew_PRL_1996}. 
Atomic coordinates were fully relaxed with the force and energy convergence criteria of 0.01\,eV/\AA\,  and 10$^{-6}$\,eV, respectively. The plane-wave cutoff energy of 520\,eV was used. The Monkhorst-pack $\Gamma$-centered $k$-point mesh of $21\times 21\times 5$ and $13\times 13\times 7$ were used for Brillouin zone sampling for single unit cells for Ti$_3$N$_2$ and Ti$_2$N, respectively. The same for TiN was taken as $11\times 11\times 11$. A Gaussian smearing of 0.05 eV width was used. The empty spheres of 1.117\AA\;   and 1.552\AA\;   Wigner-Seitz radii ($R_{\rm WS}$) were taken for Ti$_3$N$_2$ and Ti$_2$N, respectively, and positioned by examining their electron localization function and matching their centers with localized IAEs. $R_{\rm WS}$ for Ti and N are 1.217\AA\;  and 0.741\AA\;   respectively in both Ti-N MXenes.  The initial crystal structures were adopted from the Materials Project~\cite{Jain_APLM_2013}. Crystallographic presentations were made using the Vesta program~\cite{Momma_JAppliedCryst_2011}. 

The dynamics of collective excitations in a crystal called phonon are due to the vibratory movements of atoms around their atomic equilibrium positions in crystalline lattices, which help understand the local stability of the crystal.  To better understand stability of various configurations of the crystal, we performed the density functional perturbation theory (DFPT) calculations~\cite{Baroni_PRL_1987, Baroni_RevModPhys_2001}  using the Phonopy program~\cite{Togo_JPCM_2023, Togo_JPSJ_2023}  and analyzed the phonon dispersion and phonon density of states in the chosen titanium nitride compounds.  In DFPT, the system's ground state is lightly perturbed by slightly displacing atoms, and the system's responses to such perturbation are analyzed. The DFPT in the Phonopy program constructs a Hessian matrix or a ``dynamical matrix $D_{ij}(\vec{q})$,"  from the second-order derivative of the system's total energy ($E$)  with respect to atomic displacements from their equilibrium positions. Mathematically,
\begin{equation}
D_{ij}(\vec{q})=\frac{\partial^2E}{\partial x_i\partial x_j }\, 
\end{equation}
where $\vec{q}$  is the momentum transfer of phonon in reciprocal spaces,  and $x_i$ and $x_j$ are atomic displacements of $i^{th}$ and $j^{th}$ atoms from their respective equilibrium positions. 
Diagonalization of  $D_{ij}(\vec{q})$ matrix provides all the phonon modes of the crystal lattice as eigenvectors and phonon frequencies corresponding to all phonon modes of the crystal lattice as eigenvalues.   We used  $2\times 2\times 2$ supercell consisting of 96 atoms for  Ti$_2$N and  $4\times 4\times 4$ supercell containing 128 for  TiN to study phonon dynamics. The used supercells achieve phonon accuracy with vanishing imaginary vibration modes in both Ti$_2$N and TiN. The imaginary frequency does not vanish in Ti$_3$N$_2$ crystals for as big as $4\times 4\times 2$ supercell of 320 atoms. The force and energy convergence criteria of 0.01\,eV/\AA\,  and 10$^{-6}$\,eV, respectively, were used for ionic relaxation during the phonon frequency calculations. In the case of Ti$_3$N$_2$ monolayer, however, the imaginary vibration mode hardly disappears only when an as big as $6\times6\times1$ supercell is constructed and used.

\section{Elecronic structure and properties }
Studying the distribution of electrons and analyzing their behaviors in a material is crucial to understanding its electronic structure. To explore the relationship of electronic structure with their electronic properties for Ti-N crystalline systems, we theoretically calculate electronic band structure, density of states (DOS), electron localization function (ELF), work function, and charge transfer.


%
\begin{figure}[htb!]
\includegraphics[width=0.95\linewidth]{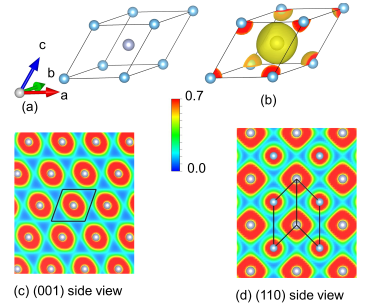}
\caption{\label{fig:TiN:ELF} ELF maps for a body-centered cubic unit cell of TiN crystal keeping N at the center of the body. Note that $\ell_{\rm FCC} = \sqrt{2}\; \ell_{\rm BCC}$. Figures show the TiN unit cell (a), $3D$ iso-surface plot for TiN with an iso-surface value of 0.6 (b), and $2D$ contour plot of ELF along (001) and (110) planes (c, d). The color scale bar shows the ELF value measured in color.}
\end{figure}
\begin{figure}[htb!]
\includegraphics[width=0.995\linewidth]{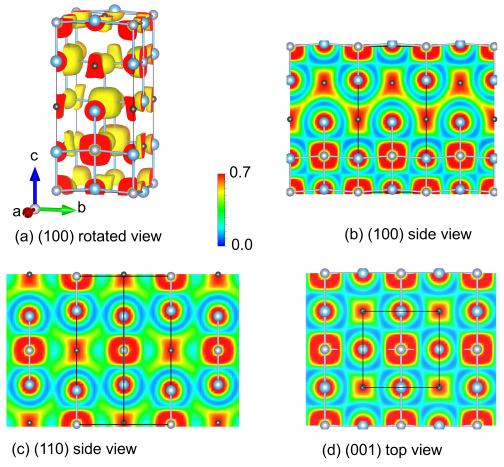}
\caption{\label{fig:Ti2N:ELF}  Rotated (100) crystal direction view of $3D$ iso-surface plot of ELF for Ti$_2$N with an isosurface value of 0.6 (a) and $2D$ contour plot of ELF along (100), (110) and (001) view for  Ti$_2$N with the same isosurface value. The color scale bar shows the ELF value measured in color. Red-colored centers in the interstitial region confirm the presence of IAEs in Ti$_2$N crystal.}
\end{figure}

\subsection{Band structure and Density of states}

Electrons' energy as a function of their wave vectors called band structure, the number of electronic states per unit electron volt of energy interval as a function of energy called DOS,  and the IAEs' contributions to the band structures and DOS at the Fermi level (E$_{\rm F}$) are calculated and presented in Figs.~\ref{fig:structure:Dos} and ~\ref{fig:structure:BNDS} for all three Ti-N crystalline systems. As expected, the band structures overlap for two different spins in the AFM configuration of the bulk Ti$_3$N$_2$. 
Each of Ti$_3$N$_2$ and Ti$_2$N has four IAEs in their crystal structures. Each IAE in Ti$_2$N receives an identical crystal lattice environment. However, IAEs in Ti$_3$N$_2$ present in different interlayer spaces and have non-identical contributions to the electronic structure. The layer of IAEs in these two interlayer spaces are denoted as A-type (strongly localized) and B-type (less localized) IAEs as shown in Fig~\ref{fig:structure}(B)(b). 
Near and at the Fermi level of Ti$_3$N$_2$, both $E_A$ and  $E_B$ empty spheres have significant contributions to the density of states. The DOS of empty spheres near and at the Fermi level is even larger than the nitrogen for both Ti$_3$N$_2$ and Ti$_2$N. The major contributions to the DOS near and above the Fermi level come from Ti in each of the titanium nitrides discussed in this paper. Orbital resolved partial density of states shows that the significant contributions to the total DOS come from the $3d$ orbital of Ti, IAEs, and $2p$ orbital of N (not shown in the figure here). At the valence band (energy band below Fermi level), the hybridization of $Ti-3d$ with empty spheres, $Ti-3d$ with $N-2p$, and $N-2p$ with empty spheres indicate the complex chemical bonding feature in Ti$_3$N$_2$ and Ti$_2$N. Since empty spheres do not contain nuclei and electrons like real atoms, to keep it simple, we did not resolve the empty sphere's basis functions into their orbitals.

\begin{figure}[htb!]
\includegraphics[width=0.975\linewidth]{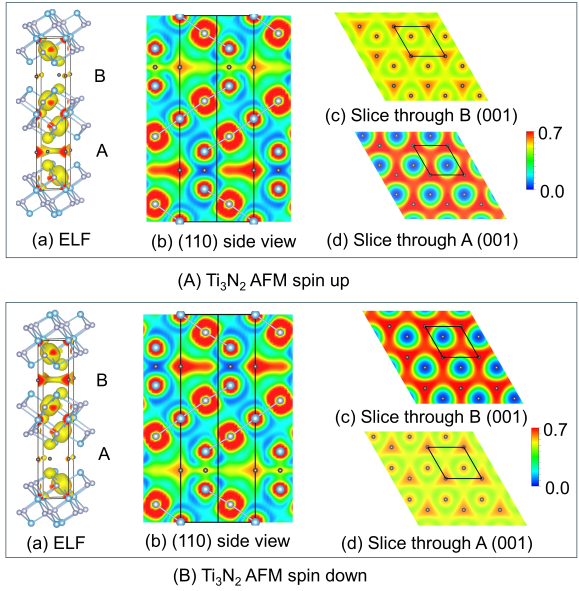}
\caption{\label{fig:Ti3N2AFM:ELF} ELF plots for Ti$_3$N$_2$  crystal in its AFM configurations for up spin (A) and down spin (B) with an iso-surface value of 0.6.  Each (A) and (B) part of the figure has a $3D$ ELF map showing in a rotated (110) view (a), a side view of $2D$ contour map of ELFs along (110) plane (b), the same along (001) plane slicing through the plane containing empty spheres $E_B$ (c) and the same for a plane containing the empty sphere $E_A$  (d). The color scale bar shows the ELF value measured in color. }
\end{figure}
\begin{figure}[htb!]
\includegraphics[width=0.975\linewidth]{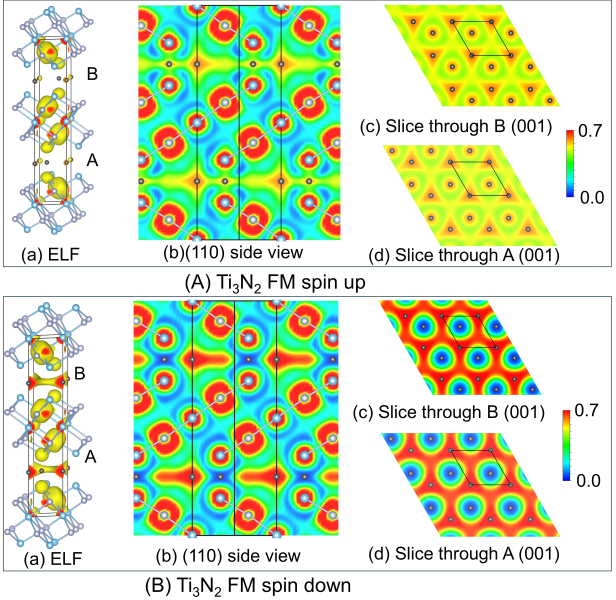}
\caption{\label{fig:Ti3N2FM:ELF} ELF plots for Ti$_3$N$_2$  crystal in its FM configurations for up spin (A) and down spin (B) with an iso-surface value of 0.6.  Each (A) and (B) part of the figure has a $3D$ ELF map showing in a rotated (110) view (a), a side view of $2D$ contour map of ELFs along (110) plane (b), the same along (001) plane slicing through the plane containing empty spheres $E_B$ (c) and the same for a plane containing the empty sphere $E_A$  (d). The color scale bar shows the ELF value measured in color. }
\end{figure}
%

%

%
\begin{figure*}[htb!]
\includegraphics[width=0.4\linewidth]{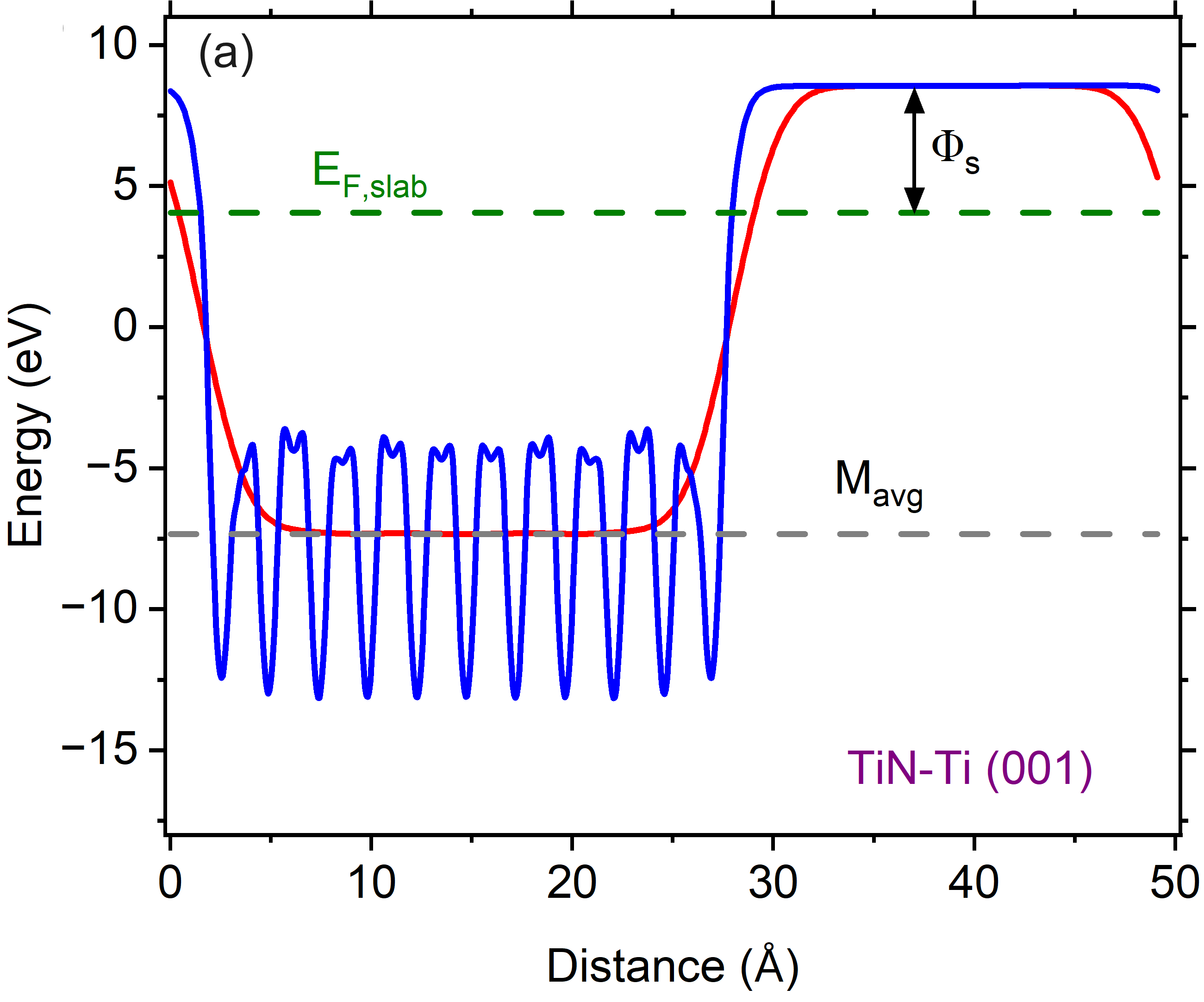}
\includegraphics[width=0.4\linewidth]{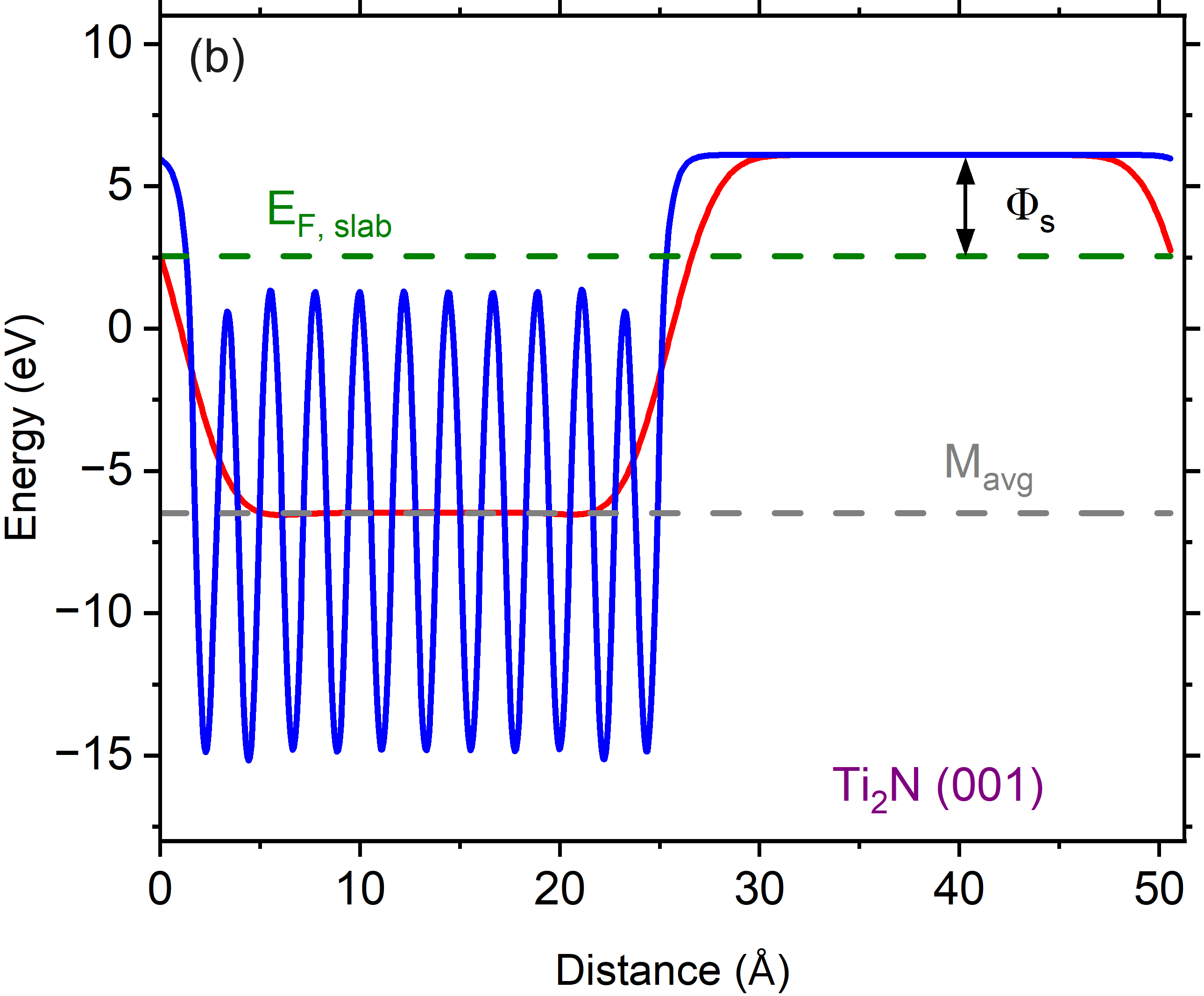}
\includegraphics[width=0.4\linewidth]{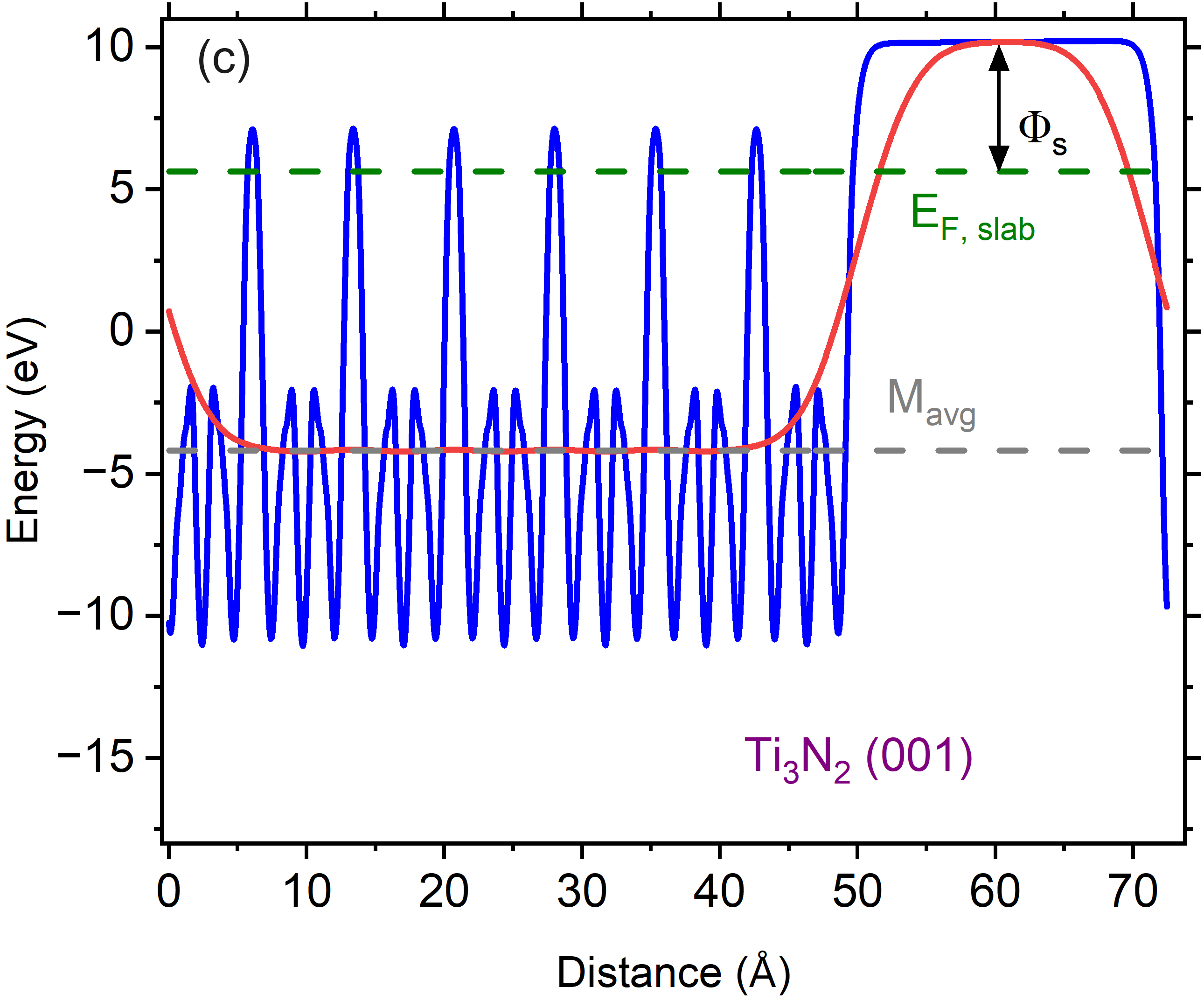}
%
\;\;
%
\includegraphics[width=0.4\linewidth]{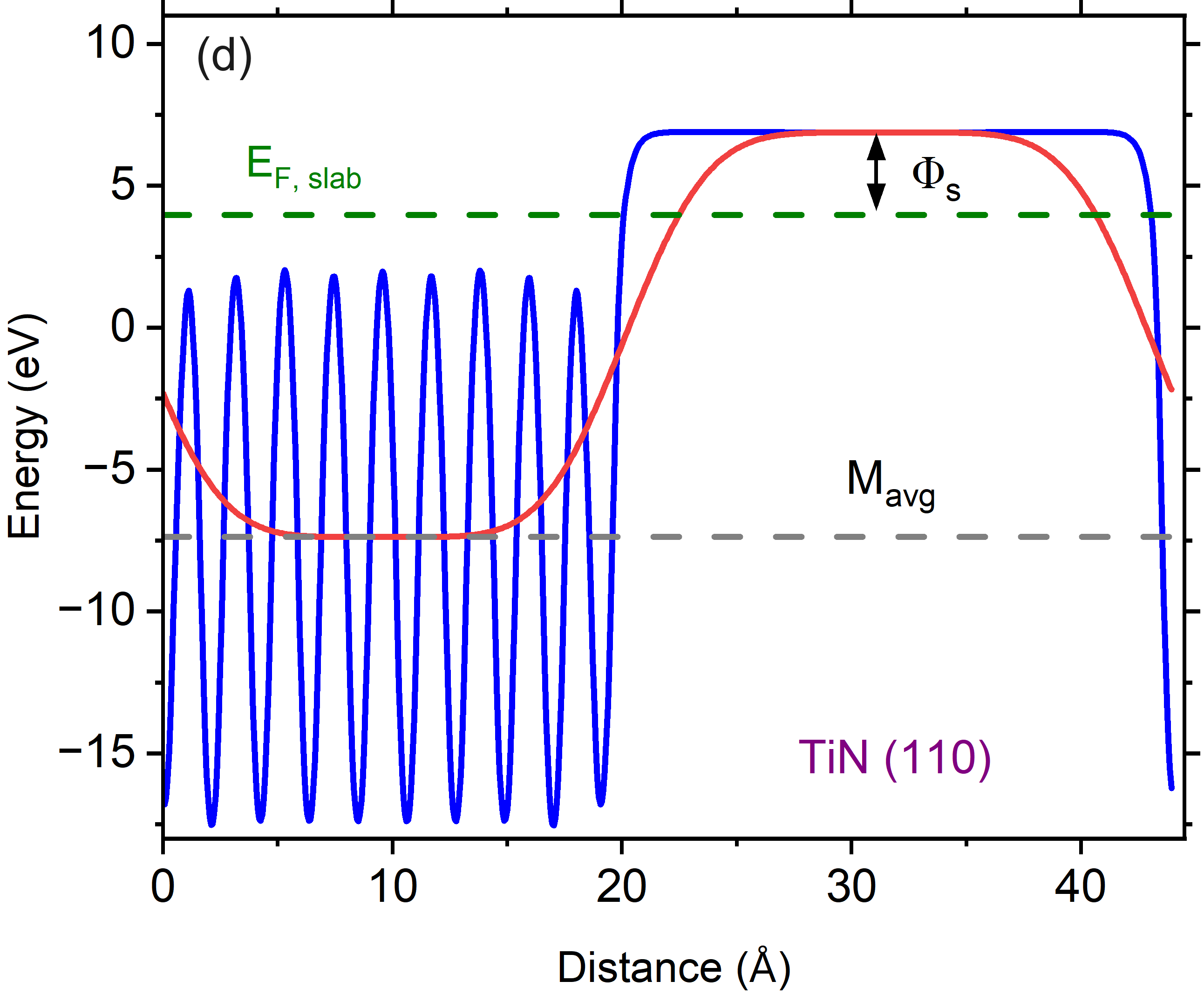}
\includegraphics[width=0.4\linewidth]{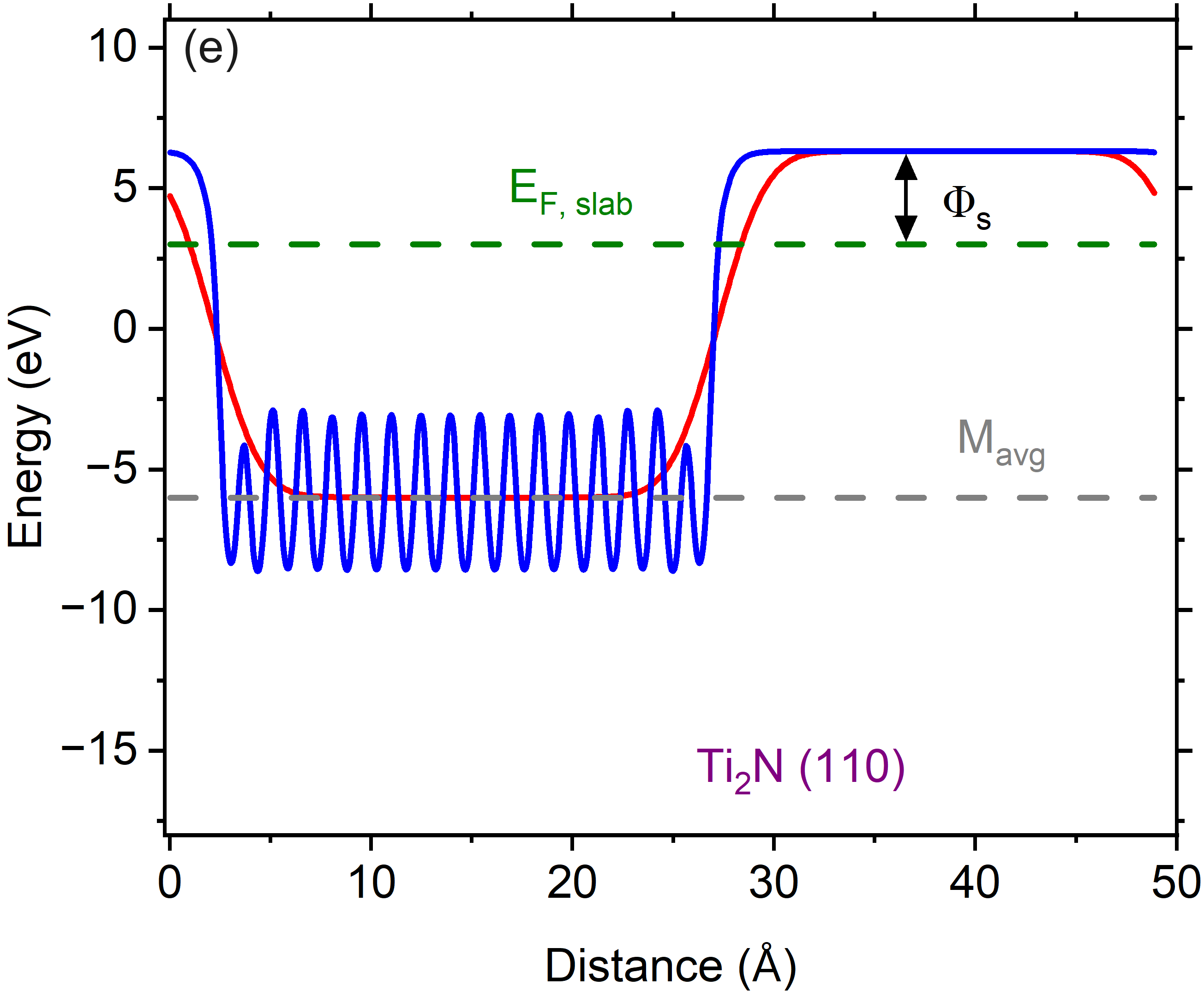}
\includegraphics[width=0.4\linewidth]{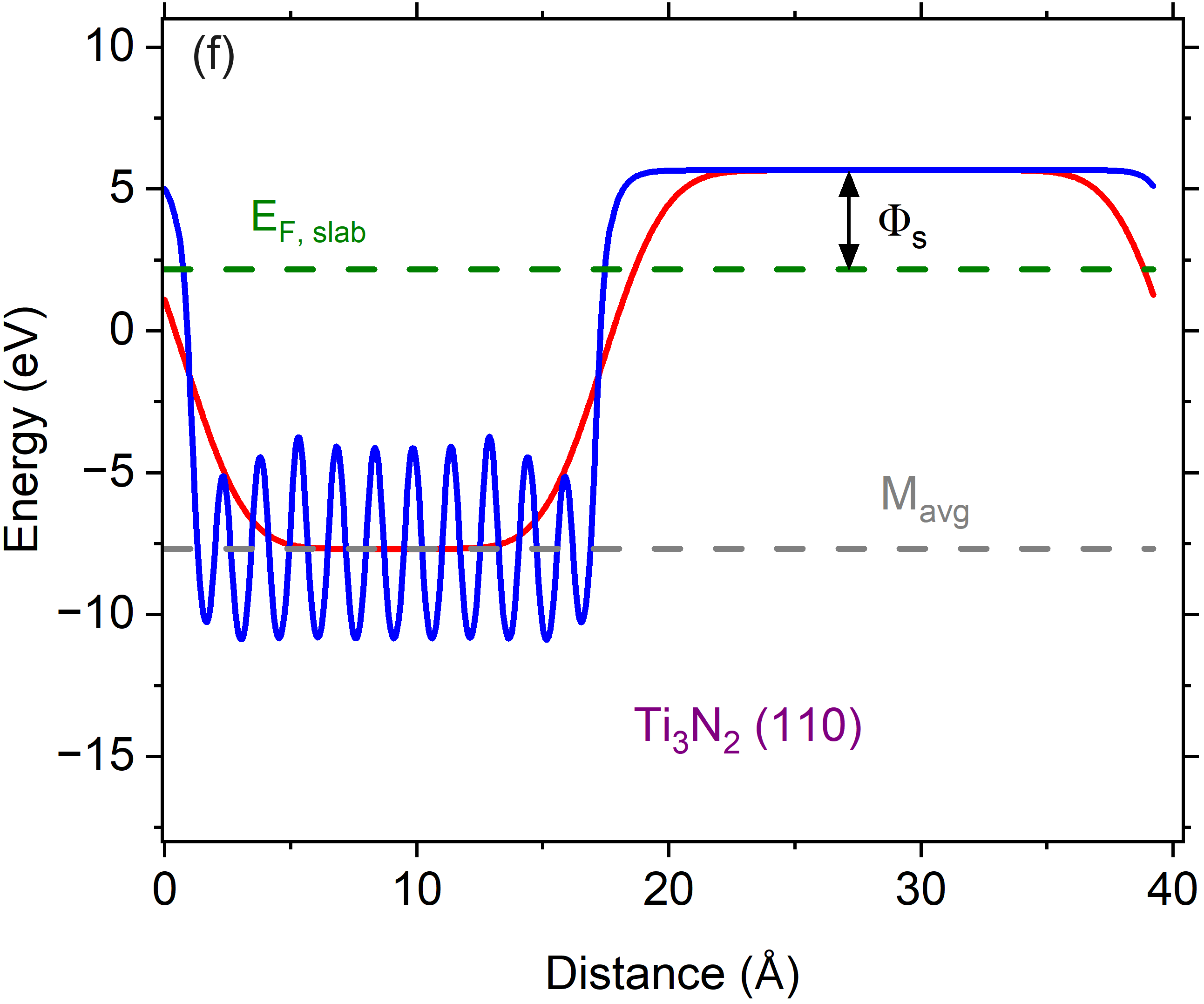}
\caption{\label{fig:structure:WF} Work function ($\Phi_S$) along (001) plane in TiN (a), Ti$_2$N (b) and AFM Ti$_3$N$_2$ (c) and the same along (110) plane in TiN (d), Ti$_2$N (e) and AFM Ti$_3$N$_2$ (f) as a function of distance. The oscillating solid blue line for each titanium nitride is the planner average energy over a plane perpendicular to the crystal axis, whereas the solid red line is the macroscopic average energy. The dashed gray line represents the average electrostatic potential, and the dashed green line is the Fermi energy of the slab E$_{\rm F,  slab}$. The same energy range is used in the vertical axis of all work function plots for better comparison. }
\end{figure*}

\subsection{Electron localization function}

To visualize and quantify electron localization in Ti-N crystals discussed in this paper, one can make use of the dimensionless  ELF given by~\cite{Becke_JChemPhys_1990, Thapa_MaterHoriz_2024}
\begin{align}
\rm{ELF} (\vec{r}) = \left[ 1 +\left\{\frac{D_\sigma(\vec{r})}{D_\sigma^0(\vec{r})}\right\}^2\right]^{-1}\,,
\end{align}
where $D_\sigma = \tau_\sigma -  (\nabla\rho_\sigma(\vec{r}))^2/(4\rho_\sigma(\vec{r})$ is the electron localization measure determining the electron pair density  curvature of an electron having identical spins at position $\vec{r}$  and $D_\sigma^0(\vec{r})$ is the same for electron gas, which reads 
\begin{align}
D_\sigma^0(\vec{r}) =\frac{3}{5} \left(6\pi^2\right)^{2/3} \rho_\sigma^{5/3}(\vec{r})\,,
\end{align}
where $\rho_\sigma(\vec{r})$ is the spin electron density at $\vec{r}$. ELF ($\vec{r}$) at any point $\vec{r}$ satisfies $0\le {\rm ELF} (\vec{r}) \le 1$ with ELF=1 indicating perfect localization. A system with a uniform charge distribution has an ELF value of 0.5. ELF is less than 0.5 if the electrons are delocalized, whereas the system with localized electrons has ELF > 0.5.

ELF maps for Ti-N crystalline systems and $2D$-contur maps along a few crystallographic directions are shown in Figs.~\ref{fig:TiN:ELF}, \ref{fig:Ti2N:ELF},  and \ref{fig:Ti3N2AFM:ELF}. To gain an accurate and deep understanding of the spatial distribution of the IAEs,  a color scale bar in the range of 0.0 to 0.7 was used with a continuous color gradient from blue (minimum ELF) to red (Maximum ELF).  In TiN, electrons are strongly localized around the Ti atom, whereas they form a uniformly dense electron gas around N. The 3D iso-surface plot of ELF and $2D$ contour plots in different directions confirm the absence of any anionic electrons in the TiN crystal. We witnessed four IAEs in a unit cell of Ti$_2$N, filled with maximally localized anionic electrons with an identical shape of iso-surface with the same color depth (see Fig.~\ref{fig:Ti2N:ELF}). Further, the (100) side view of ELF in Ti$_2$N indicates the one-dimensional distribution of IAEs within the lattice crystal where the localized IAEs are interconnected with delocalized electron channel as observed in transition metal-rich oxide and chalcogenide electride materials~\cite{Thapa_MaterHoriz_2024}. Ti$_3$N$_2$ hosts a layer of IAEs between every two adjacent pairs of MXene layers. The IAE layers in different spatial locations receive a non-identical environment, making them distinct from each other.  ELF of a Ti$_3$N$_2$ crystal depends not only on its magnetic configuration but also on its spin alignment. Anionic electrons in the Ti$_3$N$_2$  AFM configuration are comparatively more localized at $E_A$ empty spheres but less localized at $E_B$s as shown in Fig.~\ref{fig:Ti3N2AFM:ELF}. The $3D$ iso-surfaces of the localized anionic electrons make almost a dumbbell shape with slightly different weights when viewed from the (110) plane; $2D$ mapped ELF of the anionic electrons form a kite-like shape. The (001) plane (top view) ensures the two-dimensional distribution of IAEs in Ti$_3$N$_2$. $E_A$ and $E_B$ exchange the localization/delocalization feature of anionic electrons as the spin is flipped. Interestingly,  in the FM configurations of  Ti$_3$N$_2$ the anionic electrons localize in both  $E_A$ and $E_B$ spheres if electrons in the host atom species are at down spin, but these electrons delocalize if electrons in the host atom species are at up spin.

\subsection{Work function}

The work function of a solid is the energy needed to remove an electron from the solid's Fermi level to a point where the electron becomes free from its influence, such that the removed electron has no kinetic energy. The work function provides deep insights into the solid's electronic, structural, physical, and optical properties, thereby guiding the solid's manipulation, design, and optimization for various electronic, optoelectronic, nanotechnological, and photovoltaic applications. The DFT calculations for determining work function involve evaluating vacuum energy, a constant electrostatic potential value along the surface's perpendicular direction, and the Fermi level of the material and finding the difference between them. Based on the calculation route, two work function calculation methods are common. The first method evaluates a slab's  Fermi energy E$_{\rm F, slab}$ and electrostatic potential in the vacuum region $V_{\rm vac}$ using self-consistent calculations, and the work function is calculated as 
\begin{align}
\Phi_S= \rm{V}_{\rm vac}-\rm{ E}_{\rm F, slab}\,,
\end{align}
where subscript $S$ in $\Phi$ indicates that it is the first method accounting for the slab's Fermi energy. 
The second method uses a two-step calculation: \textit{(i)} computing the electrostatic potential in the vacuum region $V_{\rm vac}$ and in the slab's interior region $\rm{ V}_{\rm Int, slab}$ using a supercell model and \text{(ii)} evaluating the Fermi level of the bulk E$_{\rm F, bulk}$ from a separate calculation. Finally, $\Phi_B$ from the second method reads
\begin{align}
\Phi_B= \rm{V}_{\rm vac}-\rm{ E}_{\rm F, bulk}^{*}\,,
\end{align}
where $\rm{ E}_{\rm F, bulk}^{*}= \rm{ E}_{\rm F, bulk} +  \rm{ V}_{\rm Int, slab}$ is the effective value of bulk's Fermi level. 
For further detail, we refer to Ref.~\cite{Thapa_MaterHoriz_2024}.

\begin{table}[ht]
\small
  \caption{\; Calculated work functions for Ti-N crystals along (001) and (110) planes. TiN-Ti and TiN-N represent the specified TiN Crystal face with the Ti and N atoms termination, respectively.}
 \label{tbl:WFs}
  \begin{tabular*}{0.475\textwidth}{@{\extracolsep{\fill}}llllll}
    \hline\noalign{\smallskip}
system &plane&E$_{\rm F, bulk}^*$(eV) & E$_{\rm F, slab}$(eV) & $\Phi_S$\;(eV) & $\Phi_B$\;(eV)   \\
\hline
TiN-Ti& (001) &  4.0625   & 4.0467  & 4.5087 & 4.4929   \\
TiN-N& (001) &  2.8787   & 2.8641  & 6.7717 & 6.7571   \\
TiN& (110) &  4.0150   & 3.9755  & 2.9040 & 2.8645  \\
\hline
Ti$_2$N& (001) & 2.5569 & 2.5403 &3.5622 &3.5456 \\
{Ti$_2$N}& (110) & 3.0394 & 3.0112 & 3.3055&3.2772 \\
\hline
Ti$_3$N$_2$ &(001)&   5.6409  & 5.6373 & 4.5733  &  4.5697   \\
{Ti$_3$N$_2$}&(110)&   2.1400  & 2.1630& 3.4979& 3.5208 \\
    \hline
  \end{tabular*}
\end{table}

The calculated work functions $\Phi_S$ and $\Phi_B$ for  Ti-N systems along with the E$_{\rm F, slab}$ and $\rm{ E}_{\rm F, bulk}^{*}$ for two different planes namely (001) and (110) are presented in table~\ref{tbl:WFs}. The energy as a function of distance plots to calculate the work function along (001) plane for Ti-N compounds are shown in Fig~\ref{fig:structure:WF}.  The (001) surface of TiN is terminated with Ti or N atoms.
The calculated value of workfunctions $\Phi_S=4.5087$ and $\Phi_B=4.4929$\,eV along (001) surface with Ti atom termination are in good agreement with 4.40$-$4.53 eV measured with Kelvin Probe Force Microscopy studying surface effect and controllable aging process on effective work function of TiN films~\cite{Zhuang_AIP_2022}  and 4.30$-$4.65 eV measured on studying TiN's work function dependency on the film thickness and nitrogen flow during reactive sputter deposition~\cite{Vitale_IEEE_2011}.  The calculated $\Phi_S$ and $\Phi_B$ work functions for TiN crystal along the (001) plane are larger with N atom termination than the Ti atom termination.  The calculated work function for (110) face of TiN  $\Phi_S=2.9040$ and $\Phi_B=2.8645$\,eV are also close to  3.17\,eV calculated using a different computational package in Ref.~\cite{Calzolari_IEEE_2020}. The calculated values of the work function are as low as 3.2772\,eV along (110) face for Ti$_2$N and 3.3479\,eV along (001) face for Ti$_3$N$_2$. Being low work function MXene electrides, Ti$_2$N and Ti$_3$N$_2$ can be used as efficient electron emitters in field emission displays that have better brightness, quicker response, and wider angular view than liquid crystal displays and organic light-emitting diodes.

\begin{figure}[htb!]
\includegraphics[width=0.985\linewidth]{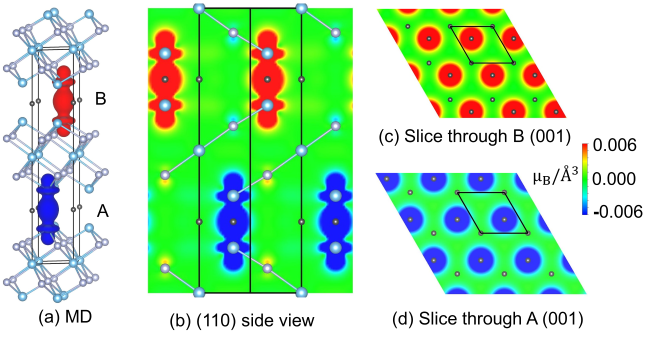}
\caption{\label{fig:Ti3N2AFM:MD} $3D$ Isosurfaces showing the magnetization density for AFM Ti$_3$N$_2$  crystal in the units of $\mu_B/{\text \AA}^3$ with isosurface value of 0.6 (a), $2D$ map of MD along (110) side view (b), and $2D$ maps of MD along (001) plane slicing through plane containing $B$ empty spheres (c)  and the plane containing $A$ empty spheres. The color scale bar shows magnetization density in the range of $-0.006$ to $0.006\; \mu_B/{\text \AA}^3$ measured in color. }
\end{figure}
\begin{figure}[htb!]
\includegraphics[width=0.985\linewidth]{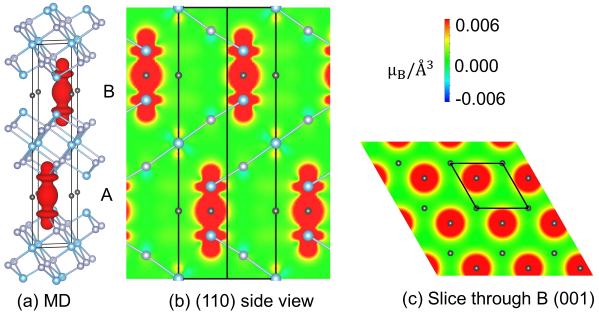}
\caption{\label{fig:Ti3N2FM:MD}  The same as in Fig.~\ref{fig:Ti3N2AFM:MD}, but for FM Ti$_3$N$_2$  crystal. }
\end{figure}
\begin{figure}[htb!]
\includegraphics[width=0.975\linewidth]{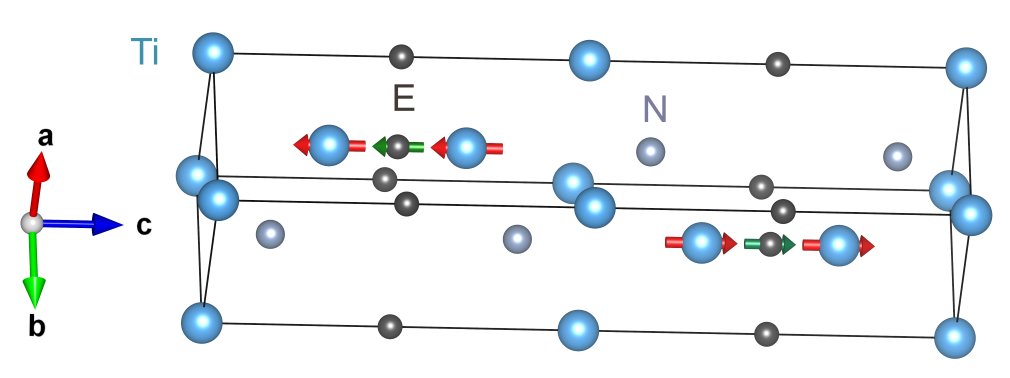}
\caption{\label{fig:structure-spin} Spin structure models of optimized Ti$_3$N$_2$ crystal.}
\end{figure}

\begin{figure}[htb!]
\includegraphics[width=0.999\linewidth]{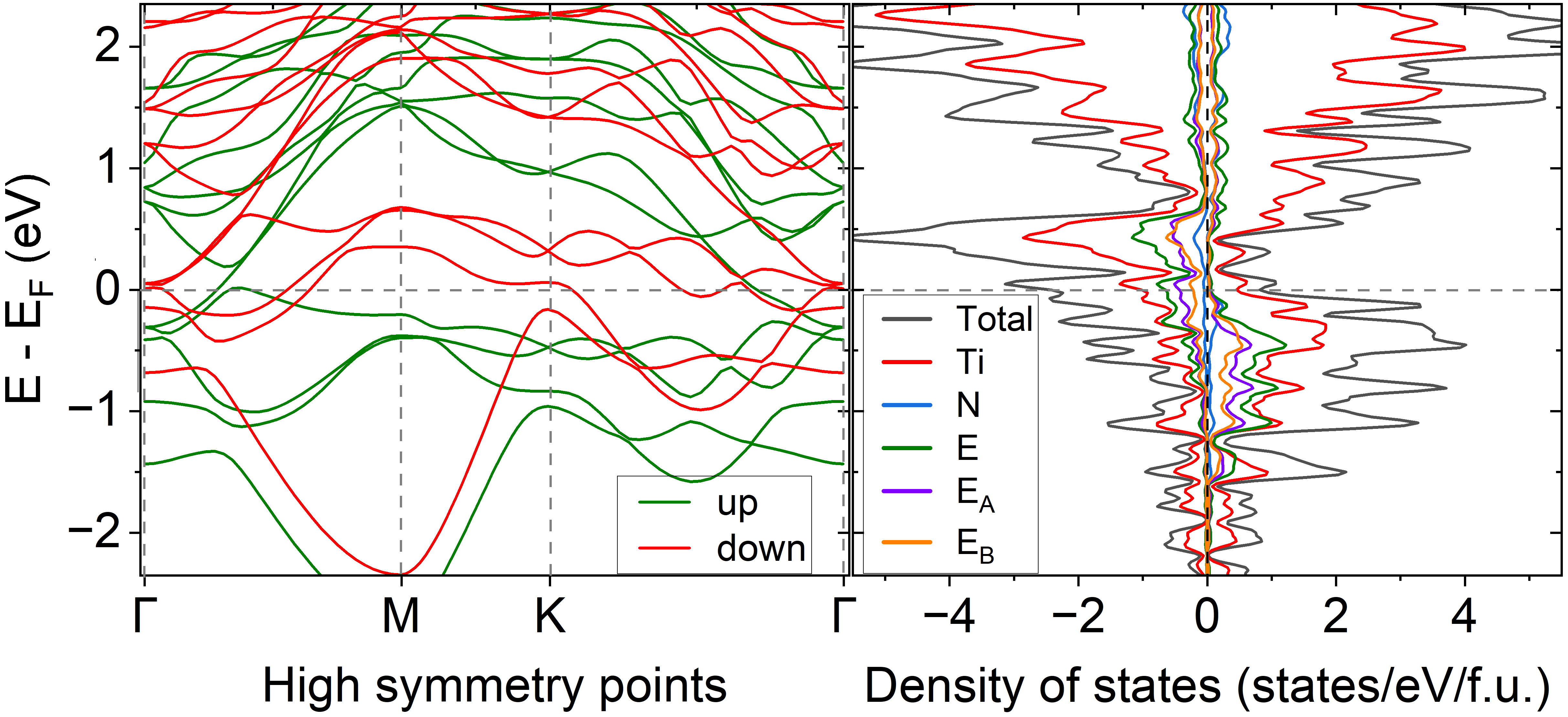}
\caption{\label{fig:bnds-dos-Ti3N2-Mono} Bandstructure of Ti$_3$N$_2$ monolayer along its high symmetry points and its atom resolved projected density of states. Fermi level (E$_{\rm F}$) is shifted to zero energy level.}
\end{figure}
\begin{figure*}[htb!]
\includegraphics[width=0.925\linewidth]{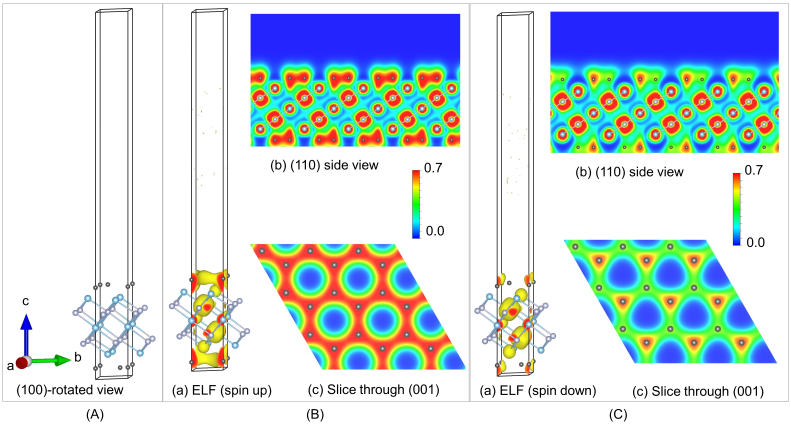}
\caption{\label{fig:Ti3N2Mono:ELF} The optimized structure of Ti$_3$N$_2$ monolayer viewing along the rotated (100) plane (A) and the ELF of Ti$_3$N$_2$ monolayer for up spin (B) and down spin(C) plotted with isosurface value of 0.6. Black color circles denote empty spheres. In each (A) and (B), $3D$ isosurfaces of ELF (a) and $2D$ contour plot mapping of the ELF viewing through (110) side plane (b) and the same taking a slice through (001) plane are presented. The color scale bar shows the ELF value measured in color with blue for minimum (0.0) and red for maximum (0.7). }
\end{figure*}
\begin{figure}[htb!]
\includegraphics[width=0.95\linewidth]{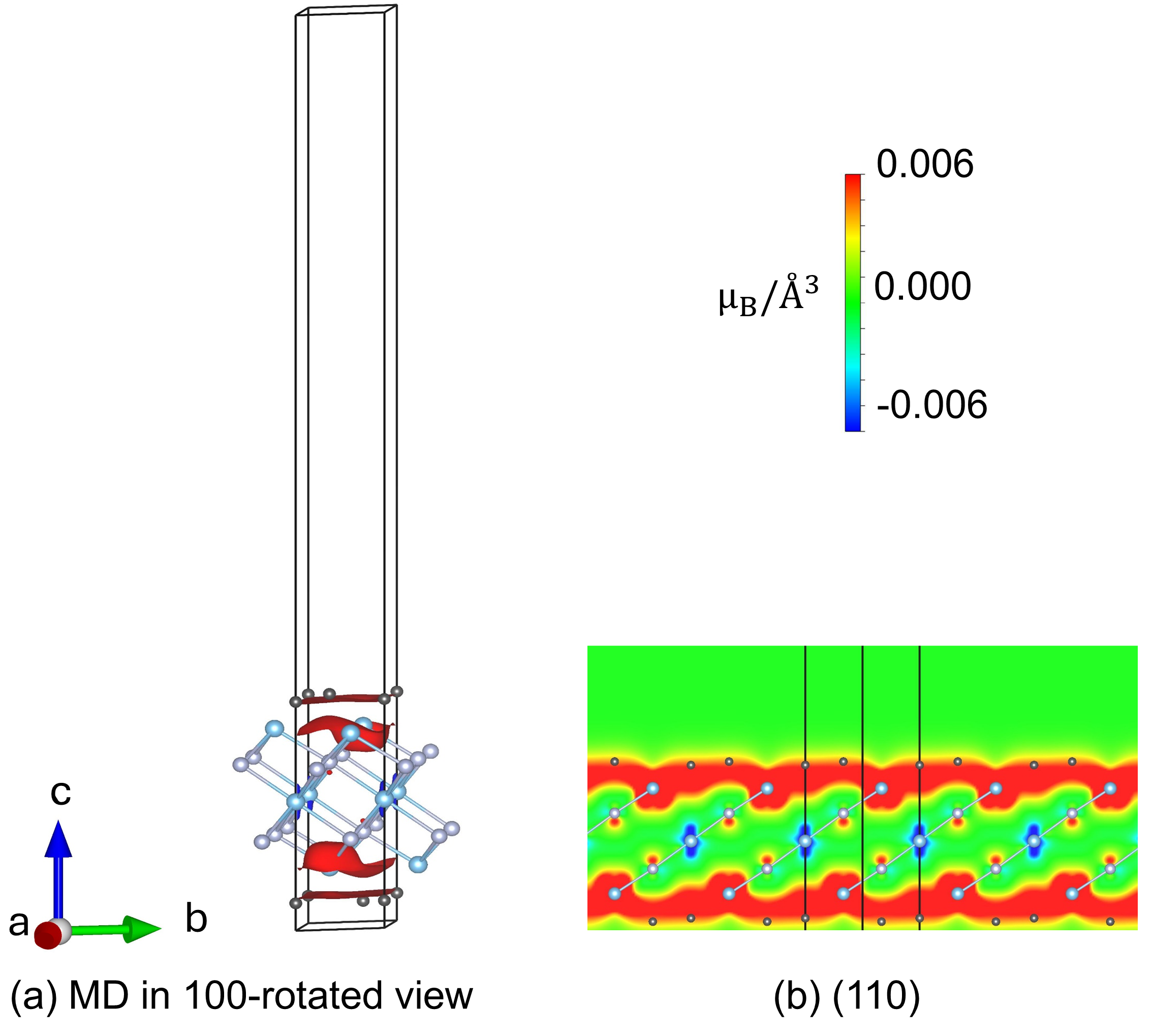}
\caption{\label{fig:Ti3N2Mono:MD} $3D$ plot of MD for Ti$_3$N$_2$ monolayer viewing along rotated (100) plane (a) and $2D$ contour maps of MD along the (110) plane. The color scale bar shows magnetization density in the range of $-0.006$ to $0.006\; \mu_B/{\text \AA}^3$ measured in color with minimum for blue and maximum for red colors.}
\end{figure}

\subsection{ Bader charge analysis}
The Bader charge analysis method~\cite{Tang_JPhysCondensMatt_2009, Sanville_JCompChem_2007} in conjunction with DFT provides a clear picture of electronic charge distribution in a solid or a molecular system, the extent of charge transfer from one atom to the other in a compound,  the oxidation state of elements in the compound, and the bonding characteristics. In non-electrides, the absence of  IAEs makes computation easy as electrons are concentrated only around the corresponding host atom species. However, the same is not true for electrides such as  Ti$_2$N and Ti$_3$N$_2$ as electrons are neither fully transferred nor shared, resulting in a positively charged lattice framework. To accurately account for the contribution of IAES in the electrides of our interest, the Bader charge analysis is performed by using total charge density with reference to the ELF, where the volume of the Bader basin around IAEs is approximated by estimating the region of the non-nuclear attractor (NNA), which can be done by locating the positions for electron density Laplacian's local maxima~\cite{Zhu_Matter_2019}. The  NNAs have localized electronic charge density that is not associated with any atomic nucleus but is significant enough to affect electride materials' physical and chemical properties.

\begin{table}[ht]
\small
  \caption{\; \label{table:OS}The average charge transfer or oxidation states on atomic species of Ti-N with reference to ELF. The plus sign denotes the charge lost, and the minus sign denotes the charge gained during the charge transfer process. The total charge is conserved. }
  \begin{tabular*}{0.48\textwidth}{@{\extracolsep{\fill}}llllll}
    \hline
compounds &Ti&N& E & E$_A$ & E$_B$ \\
\hline
TiN& +2.566 & -2.566& {}&{}&{} \\
Ti$_2$N& +2.431 & -2.318 & -2.543 &{} & {} \\
Ti$_3$N$_2$ AFM UP& +2.405 & -2.525 & -1.083  &-1.880 &-0.285\\
Ti$_3$N$_2$ AFM DN& +2.403 & -2.524 & -1.080&-1.880&-0.280  \\
Ti$_3$N$_2$ FM UP& +2.454 & -2.501 & -1.181&-1.730 &-0.633 \\
Ti$_3$N$_2$ FM DN& +2.365 & -2.540 & -2.016&-2.016 & \,0.000 \\
Mono Ti$_3$N$_2$ FM UP&   +2.422& -2.324 &	-0.654& -0.810& -0.498\\
Mono Ti$_3$N$_2$ FM DN& +2.356	& -2.658&	-0.438& -0.708& -0.169\\
\hline
  \end{tabular*}
\end{table}

The average value of charge transfer for an $i^{th}$ atomic species is calculated using $q_i=N_{VE,i}- Q_{B,i}$, where $N_{VE,i}$  is the number of pseudo-potential valence electrons and $Q_{B,i}$ is the Bader charge enclosed within the  Bader basin of the $i^{th}$  species and summarized in Table~\ref{table:OS}. The charge transfer value gives an insight into the degree of oxidation or reduction an atomic species undergoes. A positive value of charge transfer of Ti and a negative value of N and Es indicates that titanium gets oxidized while nitrogen and empty spheres get reduced. Empty spheres in the different $A$ and $B$ planes in Ti$_3$N$_2$ do not equally reduce. Interesting things happen in the FM configuration of Ti$_3$N$_2$ up spin state, both $A$ and $B$ have negative oxidation values. For the down spin state, however, only the empty spheres in the $A$ plane have negative oxidation values, while the empty spheres in the $B$ plane have an oxidation value of zero, which means they neither oxidize nor reduce, as shown in Fig. ~\ref{fig:Ti3N2FM:ELF}B.

\begin{figure*}[htb!]
\includegraphics[width=0.333\linewidth]{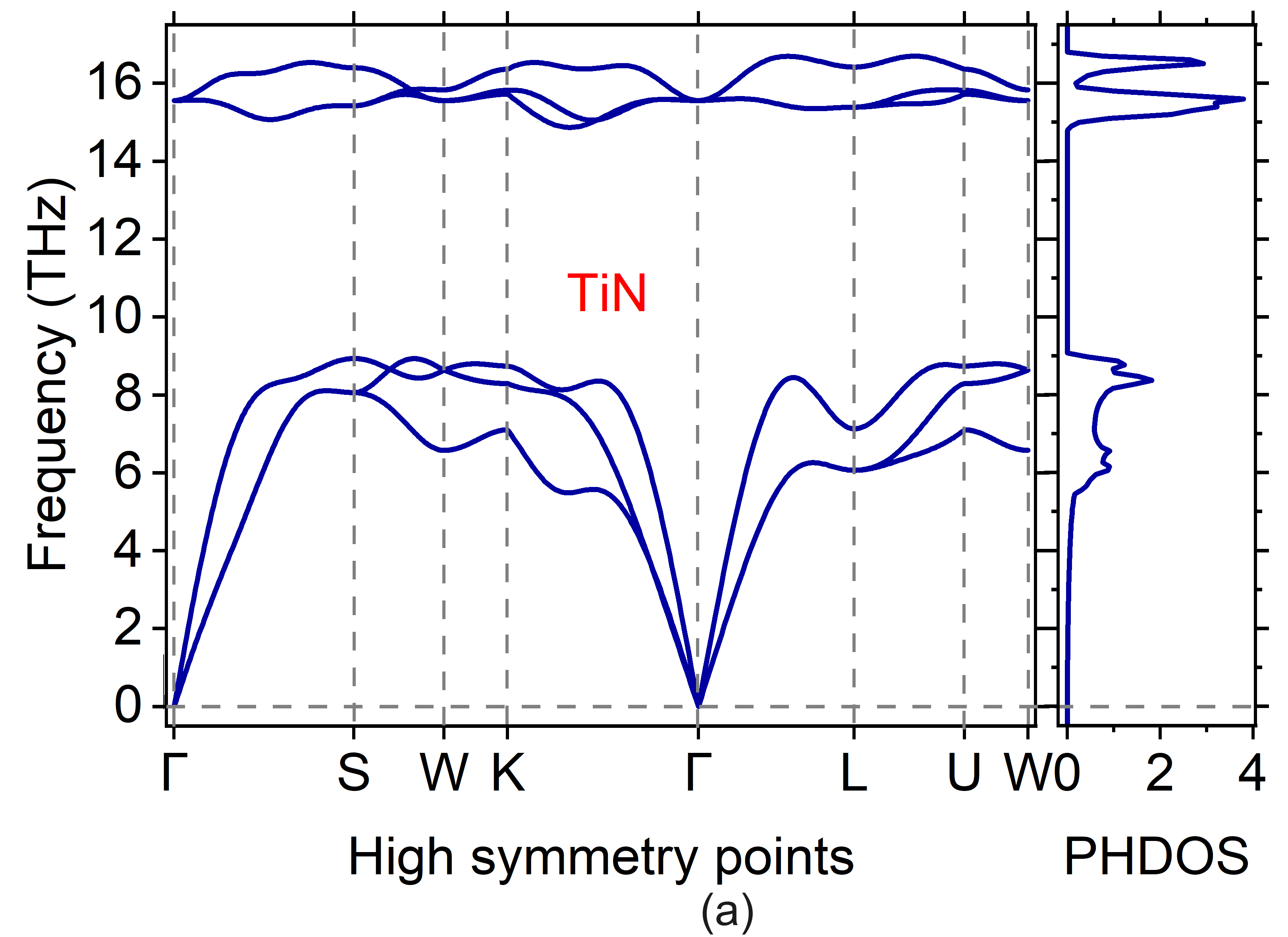}
\includegraphics[width=0.333\linewidth]{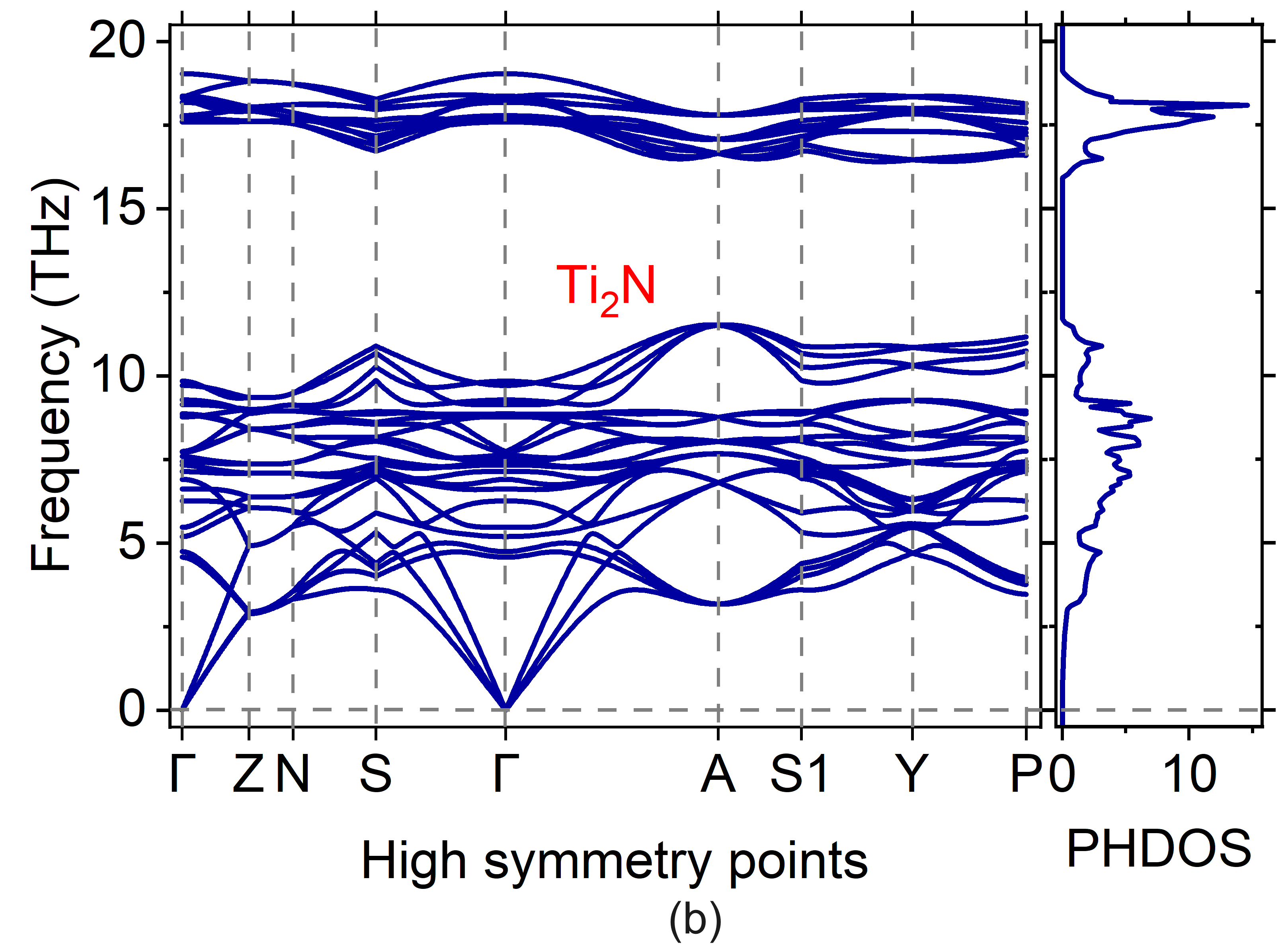}
\includegraphics[width=0.312\linewidth]{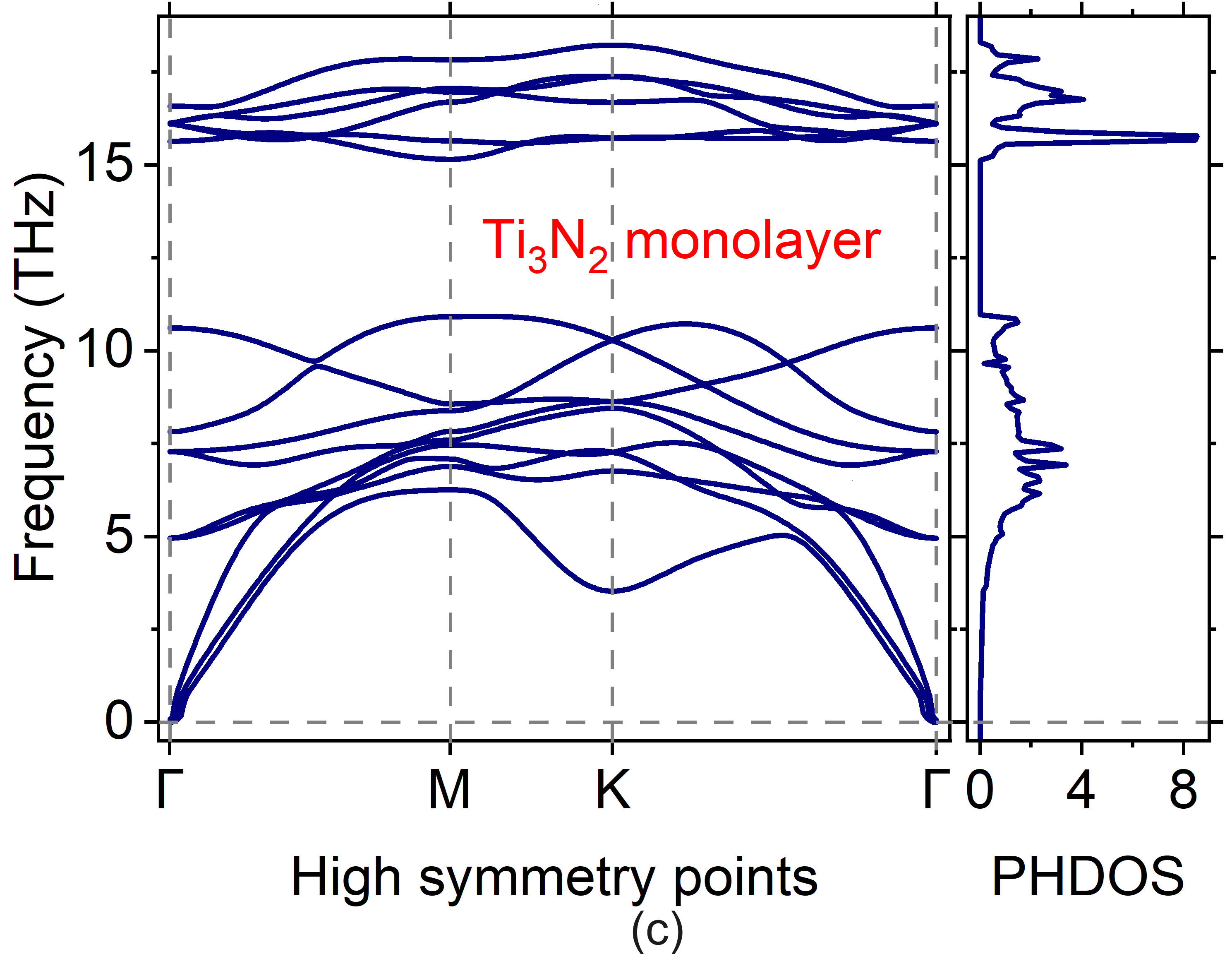}
\caption{\label{fig:phonon} Phonon dispersion curves and phonon density of states (PHDOS) for (a) TiN, (b) Ti$_2$N, and (c) a monolayer of Ti$_3$N$_2$.}
\end{figure*}

\section{Magnetism}

Nonmagnetic electrides with low work function can be excellent electron emitters, whereas magnetic electrides with low work function are excellent for magnetic tunnel junctions, and magnetic electrides with IAEs having the same spin can be used in spintronic devices for spin injection~\cite{Zhang_jacs_2023}. The spin-polarized nature of IAEs provides a step towards the correlated electronic phases in the quantum regime~\cite{Giuliani_Vignale_2005, Kim_NatMat_2022}.

Non-electride TiN and electride Ti$_2$N are non-magnetic; however, Ti$_3$N$_2$ has a complex magnetic structure. The structural analysis shows that FM and AFM states have competing stability, with the AFM configuration being slightly more stable. To understand the magnetic dipole alignment and distribution within the Ti$_3$N$_2$ crystal, we calculated the magnetization density (MD) extracted from the difference in spin up and down components of pre-converged charge density in the unit of $\mu_B/{\text \AA}^3$ for both AFM and FM configurations with an isosurface value of 0.6 and presented along with 2D contour plots of MD map along the specified plan view in Figs.~\ref{fig:Ti3N2AFM:MD} and \ref{fig:Ti3N2FM:MD}, respectively~\cite{Kim_npj_2021}. In both AFM and FM configurations, Ti$_3$N$_2$ has the magnetization density concentrated about NNA and has a vanishing value at the edge. A unit cell of Ti$_3$N$_2$ has two formula units, with 6 Ti atoms whose calculated values of magnetic moments are $-$ 0.229,  0.229,  0.229, $-$ 0.229, 0.000, and 0.000$\;\mu_B$. Nitrogens have no magnetic moment, while the four empty spheres have magnetic moments of $-$0.029, 0.029, 0.217,$-$0.217$\;\mu_B$.

Any change in the atomic coordination and local environment, symmetry,  change in electronic structure, and surface and edge effect significantly alter the magnetism of a magnetic material~\cite{Coey_1999, Freeman_PhysRev_1962, Kittel_2004, Gambardella_Science_2003}.  The MD in each configuration is concentrated near and at the IAEs, so the Ti atoms at the corners of the crystal receive a completely different electronic environment than the Ti residing inside the crystal. This leads to a complete loss of the magnetic moment for atoms at the surface and corners, while non-zero magnetism for Ti resides in the inner part of the crystal. 

The vanishing magnetic moment for Ti atoms at the edge but the unidirectional alignment of spins in the remaining atoms in each layer results in each layer in Ti$_3$N$_2$ being ferromagnetic. However, the adjacent Ti$_3$N$_2$ layers are ordered antiferromagnetically, resulting in zero magnetic moments in the bulk Ti$_3$N$_2$'s ground state. Hexagonal symmetry and interlayer antiferromagnetism indicate the possibility of altermagnetism in Ti$_3$N$_2$. However, rotation and translation operations in Ti$_3$N$_2$ cannot remove the spin degeneracy, confirming that the bulk Ti$_3$N$_2$ is only antiferromagnetic but not altermagnetic.

\section{Ti$_3$N$_2$ Monolayer}
Competing AFM and FM configurations in  Ti$_3$N$_2$ motivates a thorough study of a Ti$_3$N$_2$ single layer. The absence of inter-layer interactions, larger surface area, and electron confinement effect cause a single-layered $2D$ structure to have different physical properties than a 3D bulk counterpart. Transition metal doped Ti$_3$N$_2$ monolayers are good HER catalysts, although the nitrogen-vacancy and nitrogen-vacancy complex with most of the transition metal dopants are endothermic~\cite{Onyia_ComMatSci_2021}. Pristine and passivated  Ti$_3$N$_2$ monolayers remain metallic even after Li$^{+}$, Na$^{+}$, and Mg$^{++}$ ions adsorption with relatively low diffusion barriers providing faster charge and discharge rates if used as anode materials~\cite{Yu_Etal_RSC_2019}.  In Ref.~\cite{Aghaei_CompCond_Mater_2019}, the authors studied the impact of surface oxidation on 2D Ti$_3$N$_2$ MXene and analyzed its electron transport and optical properties and found that surface oxidation decreases metallic conductivity, absorption, and reflectivity in the visible range.  

Ti$_3$N$_2$  monolayer stabilizes in its ferromagnetic configuration and has a repeating pattern of equilateral triangles with reflections across certain axes when viewed from the $c$-axis  as shown in Fig.~\ref{fig:Ti3N2Mono:ELF}(A) and falls in hexagonal  $p\overline{3}m1$ space group with the international number of 164. The optimized  Ti$_3$N$_2$ monolayer has a lattice parameter of 2.9902\;\AA\;  in which two lattice vectors in a plane make 120$^{0}$ angle. To have an interlayer interaction-free $2D$ structure, we introduced a vacuum space, which in turn optimized with a single layer with a $c$ value of  31.496\;\AA.  Our first-principle DFT study on the electronic structure and magnetism of Ti$_3$N$_2$ monolayer indicates it to be a ferromagnetic electride, which has not yet been reported in the literature. The nominating spin-up IAEs form a two-dimensional hexatic pattern on the surfaces, whereas the spin-down IAEs are distributed in a regular triangular fashion, forming a novel Wigner crystal. 

The Ti$_3$N$_2$  monolayer is metallic, and it has a non-overlapping energy level for different spins, as shown in Fig.~\ref{fig:bnds-dos-Ti3N2-Mono}. The atomic species resolved density of states shows that the empty spheres contribute significantly to the density of states, revealing that Ti$_3$N$_2$  monolayer is also an electride. The density of states has a significantly lower value for the up spin than the down spin. Anionic electrons localization in the plane above and below the $2D$ FM Ti$_3$N$_2$ layer is larger for electrons' up spin than the down spin alignment (see Figs.~\ref{fig:Ti3N2Mono:ELF}(B) and (C)). The MD's $3D$ isosurface plot and its $2D$ contour map, as shown in Fig.~\ref{fig:Ti3N2Mono:MD}, shows that the Ti residing in the inner part of the monolayer has non-zero magnetic moment while Tis at the corner have vanishing magnetic moment, which is consistent with what we have seen in its bulk configuration. Numerically, two Ti atoms in the inner part of the monolayer have a magnetic moment of 0.594 $\mu_B$ each, vanishing magnetic moment for N atoms, and 0.170, 0.162, 0.141, and 0.135$\;\mu_B$ for four empty spheres.

\section{Phonon's dispersion and  density of states}

Using DFPT in conjunction with VASP and the Phonopy program, we calculated the phonon band structure, measuring the phonon frequency variation with their wave vector along the high symmetry point of irreducible Brillion Zone (BZ) boundaries and phonon density of states (PHDOS), showing the number of phonon states available at each frequency for TiN, Ti$_2$N, and a monolayer of Ti$_3$N$_2$ and presented in Fig.~\ref{fig:phonon}. The phonon dispersion curves show nonzero values only for the real part of vibrational frequency for TiN, Ti$_2$N, and a monolayer of Ti$_3$N$_2$, validating their dynamical stability. The DFPT calculations reveal vibrational mode instability in  Ti$_3$N$_2$ crystal for as big as $4\times 4\times 2$ supercell of 320 atoms as it has a nonzero imaginary component within the BZ.  TiN, Ti$_2$N, and a monolayer of Ti$_3$N$_2$ have frequency gaps of 5.7219, 4.2261, and 4.1545\;THz between their acoustic and optical phonon modes, which in eV reads $2.3664\times 10^{-2}$, $1.7478\times10^{-2}$, and $ 1.7181\times10^{-2}$ respectively. 


\section{Conclusions}
The electronic structure and magnetism of crystalline Ti-N systems, namely,  TiN, Ti$_2$N, and Ti$_3$N$_2$, are studied using the first principle DFT calculating various physical quantities such as formation enthalpy, cohesive energy, exfoliation energy to extract layered MXene structure from Al-based MAX phase, band structure, DOS, charge transfer, work function, ELF, magnetic moment, and MD. Producing the results consistent with the available results for well-studied crystalline TiN systems and extending a detailed analysis to Ti$_2$N and Ti$_3$N$_2$ MXene, we discover a new family of electride materials having low work functions at least along some crystallographic direction and variety in magnetic properties. TiN, which does not follow the general formula of MXene, is neither a  magnetic nor an electride.  Ti$_2$N is a non-magnetic MXene electride. Bulk Ti$_3$N$_2$ Mxene has competing AFM and FM configurations with a slight preference for AFM configuration, which is a magnetic electride for both configurations. Calculations show a monolayer of Ti$_3$N$_2$ MXene is a ferromagnetic electride. 



\section*{Acknowledgements}
The Department of Energy BES-RENEW award number DE-SC0024611 has supported this research work. 
D. T. and S. K. acknowledge the Center for Computational Assisted Science and Technology (CCAST) at North Dakota State University for providing computational resources for this study. 
.





\bibliography{arxiv_rscA} 
\bibliographystyle{rsc} 

\end{document}